\documentclass[11pt]{article}
\usepackage{amsmath, amssymb, geometry, graphicx}
\usepackage{hyperref}
\usepackage{physics}
\usepackage{bm}
\usepackage[numbers]{natbib}
\usepackage{float}
\usepackage[font=small,skip=10pt]{caption}
\usepackage{subcaption}  
\usepackage{graphicx}
\usepackage{caption}
\usepackage{xcolor}

\title{Integrating Score-Based Generative
Modeling and Neural ODEs for
Accurate Representation of
Multiscale Chaotic Dynamics}
\author{Giulio Del Felice\thanks{Email: giuliodf@mit.edu | gdelfelice@ethz.ch} \\ Department of Physics, Eidgenössische Technische Hochschule Zürich, \\ Zürich 8092,
Switzerland, \\ Department of Mathematics, Massachusetts Institute of Technology \\
\and Ludovico Theo Giorgini\thanks{Email: ludogio@mit.edu} \\
Department of Mathematics, Massachusetts Institute of Technology, \\
Cambridge, MA 02139, USA}
\date{}

\begin{document}
\maketitle

\begin{abstract}

Multiscale dynamical systems characterized by interacting fast and slow processes are ubiquitous across scientific domains, from climate dynamics to fluid mechanics. Accurate modeling of such systems requires capturing both the long-term statistical properties governed by slow variables and the short-term transient dynamics driven by fast chaotic processes. We present a hybrid data-driven framework that integrates score-based generative modeling with Neural Ordinary Differential Equations (NODEs) to construct reduced-order models (ROMs) capable of reproducing both regimes. The slow dynamics are represented by a Langevin equation whose drift is informed by a score function learned via the K-means Gaussian Mixture Model (KGMM) method, ensuring faithful reproduction of the system's invariant measure. The fast chaotic forcing is modeled by a NODE trained on delay-embedded residuals extracted from observed trajectories, replacing conventional Gaussian noise approximations. We validate this approach on a hierarchy of prototypical metastable systems driven by Lorenz 63 dynamics, including bistable potentials with additive and multiplicative forcing, and tristable non-autonomous systems with cycloperiodic components. Our results demonstrate that the hybrid framework maintains statistical consistency over long time horizons while accurately forecasting rare critical transitions between metastable states with lead times approaching the Lyapunov time of the chaotic driver. This work establishes a principled methodology for combining statistical closure techniques with explicit surrogate models of fast dynamics, offering a pathway toward predictive modeling of complex multiscale phenomena where both long-term statistics and short-term transients are essential.

\end{abstract}

\section{Introduction}
\label{sec:introduction}

Multiscale complex systems pervade numerous scientific domains, from climate science \cite{Hasselmann1976, LucariniChekroun2023, majda_fdt, keyes2023stochastic, giorgini2022non} and fluid dynamics \cite{steinhauser2017computational} to biological processes \cite{meier2009multiscale} and financial markets \cite{in2013introduction}. These systems are characterized by interacting processes that evolve across widely disparate temporal and spatial scales, posing significant challenges for computational modeling. However, in many cases, most of the variance in observed data can be explained by a few dominant slow modes, motivating the development of reduced-order models (ROMs) that focus exclusively on these slow variables. Projection-based reduction techniques, exemplified by the Mori--Zwanzig formalism \cite{Zwanzig1961Memory, Mori1965Transport}, reveal that eliminating fast degrees of freedom yields an effective description with two key ingredients: a memory kernel encoding non-Markovian effects, and a fluctuating force representing the influence of unresolved fast processes. To incorporate these effects into tractable ROMs, two main strategies have emerged: (i) augmenting the dimension of the slow variables to explicitly account for memory—an approach that in principle requires infinite dimensions but can be approximated by finite-dimensional embeddings when memory is sufficiently short or structured \cite{Lei2016GLE, Vroylandt2022LLnonMarkov, Ge2024PRL}; or (ii) introducing a statistical closure that modifies the drift term of the reduced dynamics to faithfully reproduce the observed statistical properties of the slow variables without dimensional augmentation \cite{pavliotis2008multiscale, chorin2015discrete}. This approach is particularly valuable in climate science, where full-scale simulations remain computationally intractable despite ongoing advances in computing infrastructure \cite{change2007climate, schneider2017earth}. In such closures, the fast timescales are typically modeled as Gaussian colored noise processes. This approximation is often justified by the central limit theorem: when a large number of weakly dependent fast processes contribute collectively to the dynamics at small timescales, their combined effect can be well approximated by Gaussian colored stochastic forcing \cite{UhlenbeckOrnstein1930, JungHanggi1987UCNA, Berner2017StochParam}.

A critical limitation of many data-driven modeling approaches is that reconstructing drift terms exclusively from short-term dynamical trajectories often yields models that fail to preserve the system's invariant measure—the probability distribution reflecting its equilibrium or statistically steady state \cite{platt2023constraining, callaham2022empirical}. This deficiency arises because optimizing solely for short-horizon prediction accuracy does not inherently ensure correct reproduction of long-term statistical properties, such as probability density functions (PDFs), autocorrelation functions (ACFs), and response to external perturbations. Recent advances in data-driven modeling methods have demonstrated that the statistical properties of multiscale systems—including the invariant measure and linear response—can be estimated more reliably from coarse-grained observational data and steady-state statistics than from fine-grained short-time trajectories \cite{falasca2024data, falasca2025FDT, giorgini2025learning, giorgini2024reduced, souza2024representing_a, souza2024representing_b, souza2024modified}.

A fundamental methodological advance for statistically faithful ROMs has emerged from recent developments in score-based generative modeling. Score-matching techniques \cite{Hyvarinen2005Score, Vincent2011DAE} and modern diffusion models \cite{SongErmon2019NCSN, Ho2020DDPM, song_sde} provide a principled route to estimating the score function—the gradient of the logarithm of the steady-state distribution—directly from high-dimensional, nonlinear observational data without requiring explicit knowledge of the normalizing constant. The availability of learned score functions has enabled fundamental breakthroughs across statistical physics and climate science: from data-driven construction of response functions and fluctuation--dissipation relations \cite{marconi2008, giorgini_response_theory, cooper2011climate, baldovin2020understanding, ghil2020physics, giorgini2025predicting, Kubo1966FDT, Ruelle1998GLRF, Ruelle2009Review}, to statistical parameter calibration \cite{giorgini2025statistical}, and ultimately to the synthesis of reduced-order models \cite{giorgini2025data, giorgini2025reduced, giorgini2025kgmm}. In these score-informed ROMs, knowledge of the score function enables a data-driven estimation of the statistical closure mentioned above: by embedding the learned score into the drift term of a Langevin equation for the slow variables, one constructs a ROM that accurately captures both the observed steady-state distribution and the temporal autocorrelations, thereby preserving the system's invariant measure by construction.

While Gaussian colored-noise surrogates for fast processes are computationally efficient and often adequate when the system exhibits well-separated timescales and frequent exploration of the state space, they prove insufficient in scenarios where accurate prediction of fast transients is essential for capturing abrupt changes in the slow variables. This limitation is particularly acute in metastable systems, which possess multiple long-lived quasi-stable states separated by energy barriers. In such systems, transitions between metastable regimes are rare events often triggered by intermittent bursts, coherent structures, or other non-Gaussian features of the fast dynamics \cite{Kramers1940Physica, Hanggi1990Kramers, lim2020predicting, giorgini2020precursors}. Accurately predicting these rare transitions—critical for understanding phenomena ranging from climate tipping points \cite{LucariniChekroun2023} to seismic stick-slip events \cite{Fenichel1979GSPT}—requires moving beyond the Gaussian-noise approximation and explicitly modeling the fast dynamics with sufficient fidelity to capture the intermittency, memory, and state-dependent amplification that govern escape from metastable wells.

In this work, we propose a hybrid reduced-order modeling framework that integrates score-based statistical closures with explicit surrogate models for fast chaotic drivers. Specifically, we model the fast process using a Neural Ordinary Differential Equation (NODE) \cite{Chen2018Neural, Rackauckas2020UDE, giorgini2021modeling}, a continuous-time universal function approximator that parametrizes the right-hand side of an ODE with a neural network. NODEs have demonstrated remarkable success in capturing the complex, often chaotic trajectories of nonlinear dynamical systems while remaining differentiable and amenable to gradient-based training. We leverage the score-informed ROM framework developed in \cite{giorgini2025data, giorgini2025reduced}, which describes the slow variable via a Langevin equation whose drift term is constructed from the learned score function, ensuring correct steady-state statistics by design. To this score-driven slow dynamics, we couple a NODE-based surrogate for the high-dimensional chaotic fast process, replacing the standard Gaussian colored-noise assumption. We apply and validate this hybrid methodology on a suite of prototypical one-dimensional multistable systems driven by the chaotic Lorenz 63 attractor \cite{lorenz2017deterministic}, exploring several challenging scenarios including cyclostationary systems and systems with state-dependent amplification of the fast forcing. While similar metastable systems with chaotic fast forcing have been studied previously \cite{lim2020predicting}, those approaches typically assume knowledge of the analytic form of the slow-variable dynamics; in contrast, our framework is entirely data-driven, learning both the statistical closure and the fast-dynamics surrogate directly from observations without requiring explicit knowledge of the governing equations. This approach aims to preserve both the long-time statistical fidelity afforded by score-based closures and the short-time dynamical accuracy required to resolve rare metastable transitions.

The remainder of this manuscript is organized as follows. Section~\ref{sec:methods} develops the mathematical framework, beginning with a derivation of the generalized Langevin equation for the slow variable via timescale separation, followed by the construction of score-based statistical closures using the KGMM method, and concluding with the delay-embedding and NODE-based approach for learning the fast chaotic dynamics. Section~\ref{sec:experiments} presents comprehensive numerical experiments on four benchmark systems: a bistable system with additive Lorenz 63 forcing (Example~I), a bistable system with state-dependent multiplicative forcing (Example~II), and a tristable non-autonomous system with cycloperiodic forcing treated in both slow-timescale (Example~IIIa) and fast-timescale (Example~IIIb) regimes. For each system, we assess both long-term statistical consistency and short-term forecasting accuracy, including the ability to anticipate critical transitions between metastable states. Section~\ref{sec:conclusions} summarizes our findings, discusses the implications of the hybrid modeling approach, and outlines directions for future work on higher-dimensional real-world systems. Technical details, including the derivation of the effective white-noise approximation and supplementary NODE performance metrics, are provided in the Appendices.

\section{Methods}
\label{sec:methods}
We investigate a dynamical system characterized by the interaction between a slow variable \(\vec{x} \in \mathbb{R}^{D_x}\) and a fast variable \(\vec{y} \in \mathbb{R}^{D_y}\). In our framework, only the variable \(\vec{x}\) is experimentally accessible through measurements. The temporal evolution of this system is governed by the following coupled differential equations:
\begin{align}
    \dot{\vec{x}} &= \vec{f}_x(\vec{x}, \vec{y}), \\
    \dot{\vec{y}} &= \vec{f}_y(\vec{x}, \vec{y}),
\end{align}
where \(\vec{f}_x: \mathbb{R}^{D_x} \times \mathbb{R}^{D_y} \to \mathbb{R}^{D_x}\) and \(\vec{f}_y: \mathbb{R}^{D_x} \times \mathbb{R}^{D_y} \to \mathbb{R}^{D_y}\) represent the respective deterministic force fields acting on each variable.

\subsection{Dynamics of the Slow Variable}
\label{subsec:slow_variable_dynamics}

To derive the effective dynamics of the slow variable \(\vec{x}\), we employ a systematic approximation based on timescale separation. We decompose the fast variable \(\vec{y}\) into its conditional average value and fluctuations around this average:
\begin{equation}
    \vec{y} = \langle \vec{y} \mid \vec{x} \rangle + \vec{y}',
\end{equation}
where \(\langle \vec{y} \mid \vec{x} \rangle\) represents the average value of \(\vec{y}\) conditioned on \(\vec{x}\), and \(\vec{y}'\) captures the rapid fluctuations. Assuming that these fluctuations are small, we can perform a first-order Taylor expansion of \(\vec{f}_x(\vec{x}, \vec{y})\) around \(\langle \vec{y} \mid \vec{x} \rangle\):
\begin{equation}
    \dot{\vec{x}} \approx \vec{f}_x(\vec{x},\langle \vec{y} \mid \vec{x} \rangle) + \mathbf{J}_{\vec{y}}\vec{f}_x(\vec{x},\langle \vec{y} \mid \vec{x} \rangle)\vec{y}',
    \label{eq:x-linearized}
\end{equation}
where \(\mathbf{J}_{\vec{y}}\vec{f}_x = \frac{\partial \vec{f}_x}{\partial \vec{y}}\) is the $D_x \times D_y$ Jacobian matrix representing how changes in the fast variable affect the dynamics of the slow variable.

By definition $\vec y'=\vec y-\langle\vec y\mid\vec x\rangle$, so differentiating and using the chain rule gives
\begin{equation}
    \dot{\vec y}'(t)
    \;=\;
    \vec f_y\!\big(\vec x(t),\langle\vec y\mid \vec x(t)\rangle+\vec y'(t)\big)
    \;-\;
    \mathbf{D}_{\vec x}\langle\vec y\mid \vec x\rangle\big|_{\vec x(t)}\,\dot{\vec x}(t),
\end{equation}
with $\mathbf{D}_{\vec x}\langle\vec y\mid \vec x\rangle$ the Jacobian of the conditional mean.
Linearizing in $\vec y'$ yields
\begin{equation}
    \dot{\vec y}'(t)
    \;\approx\;
    \underbrace{\mathbf{A}(\vec x(t))}_{\displaystyle \mathbf{J}_{\vec y}\vec f_y(\vec x,\langle\vec y\mid\vec x\rangle)}\,\vec y'(t)
    \;-\;
    \underbrace{\mathbf{B}(\vec x(t))}_{\displaystyle \mathbf{D}_{\vec x}\langle\vec y\mid \vec x\rangle}\,\dot{\vec x}(t)
    \;+\;
    \vec{\eta}(t),
    \label{eq:yprime-linear}
\end{equation}
where $\vec{\eta}(t)$ collects the (fast) inhomogeneous part $\vec f_y(\vec x,\langle\vec y\mid\vec x\rangle)$, any explicit fast stochastic forcing, and higher-order terms neglected in the linearization.

Let $\mathbf{G}(t,s;\vec x)$ solve $\partial_t \mathbf{G}(t,s)=\mathbf{A}(\vec x(t))\,\mathbf{G}(t,s)$ with $\mathbf{G}(s,s)=\mathbf{I}$. The solution of \eqref{eq:yprime-linear} is
\begin{equation}
\begin{aligned}
    \vec y'(t)
    \;=\;
    &\;\mathbf{G}(t,t_0;\vec x)\,\vec y'(t_0)
    \;-\;\int_{t_0}^{t}\!\mathbf{G}(t,s;\vec x)\,\mathbf{B}\big(\vec x(s)\big)\,\dot{\vec x}(s)\,ds \\
    &\;+\;\int_{t_0}^{t}\!\mathbf{G}(t,s;\vec x)\,\vec{\eta}(s)\,ds.
\end{aligned}
\label{eq:yprime-solution}
\end{equation}

Insert \eqref{eq:yprime-solution} into \eqref{eq:x-linearized}:
\begin{equation}
\begin{aligned}
    \dot{\vec x}(t)
    \;\approx\;
    &\;\vec f_x\!\big(\vec x(t),\langle\vec y\mid \vec x(t)\rangle\big) \\
    &\;-\;\int_{t_0}^{t}\!\underbrace{\mathbf{J}_{\vec y}\vec f_x\!\big(\vec x(t),\langle\vec y\mid \vec x(t)\rangle\big)\,
    \mathbf{G}(t,s;\vec x)\,\mathbf{B}\!\big(\vec x(s)\big)}_{\displaystyle \mathbf{K}\big(\vec x(t),\vec x(s);t-s\big)}\,
    \dot{\vec x}(s)\,ds \\
    &\;+\;\underbrace{\mathbf{J}_{\vec y}\vec f_x\!\big(\vec x(t),\langle\vec y\mid \vec x(t)\rangle\big)
    \Big[
        \mathbf{G}(t,t_0;\vec x)\,\vec y'(t_0)
        + \int_{t_0}^{t}\!\mathbf{G}(t,s;\vec x)\,\vec{\eta}(s)\,ds
    \Big]}_{\displaystyle \vec{\zeta}(\vec{x},t)}.
\end{aligned}
\label{eq:gle}
\end{equation}
Equation \eqref{eq:gle} is a \emph{generalized Langevin equation (GLE)} for $\vec x$ with a matrix memory kernel $\mathbf{K}$ and a colored random force $\vec{\zeta}(\vec{x},t)$. Specifically, we characterize $\vec{\zeta}(\vec{x},t)$ by zero mean and exponentially decaying temporal correlations:
\begin{align}
    \langle \vec{\zeta}(\vec{x},t) \rangle &= \vec{0}, \\
    \langle \vec{\zeta}(\vec{x},t) \vec{\zeta}^{\top}(\vec{x},s) \rangle &\approx \bm{D}_{0}(\vec{x}) \odot e^{-|t-s|/\tau},
\end{align}
where $\bm{D}_{0}(\vec{x}) \in \mathbb{R}^{D_x \times D_x}$ is the position-dependent covariance (or diffusion) matrix of the effective random force, $\tau$ is a single characteristic decorrelation time representing the typical relaxation timescale of the colored forcing, and $\odot$ denotes element-wise multiplication.

To make this more explicit, we factorize the colored random force as
\begin{equation}
\vec{\zeta}(\vec{x},t) = \bm{\Sigma}_0(\vec{x})\vec{\eta}(t),
\label{eq:zeta_factorization}
\end{equation}
where $\vec{\eta}(t)$ is a colored noise process with zero mean and unit variance, satisfying $\langle \eta_i(t)\eta_j(s) \rangle = \delta_{ij} e^{-|t-s|/\tau}$, and $\bm{\Sigma}_0(\vec{x})$ is obtained from the Cholesky decomposition of $\bm{D}_0(\vec{x})$:
\begin{equation}
\bm{\Sigma}_0(\vec{x})\bm{\Sigma}_0^{\top}(\vec{x}) = \bm{D}_0(\vec{x}).
\end{equation}

For observations on time scales significantly longer than the decorrelation time $\tau$, the colored noise can be approximated by an effective white noise process. Following the rigorous derivation in Appendix~\ref{sec:appendix_white_noise}, we define the effective diffusion matrix
\begin{equation}
\bm{D}(\vec{x}) = \tau \bm{D}_0(\vec{x}),
\end{equation}
and the corresponding effective white-noise amplitude matrix
\begin{equation}
\bm{\Sigma}(\vec{x}) = \sqrt{\tau}\,\bm{\Sigma}_0(\vec{x}) = \text{chol}\big(\bm{D}(\vec{x})\big).
\label{eq:white_noise_amplitude}
\end{equation}
This transformation preserves the statistical properties of the process when the observation time scale greatly exceeds $\tau$, replacing the colored noise $\bm{\Sigma}_0(\vec{x})\vec{\eta}(t)$ with white noise $\bm{\Sigma}(\vec{x})\vec{\xi}(t)$, where $\vec{\xi}(t)$ is standard white noise with $\langle \xi_i(t)\xi_j(s) \rangle = \delta_{ij}\delta(t-s)$. The scaling factor $\sqrt{\tau}$ arises from matching the time-integrated covariance of the exponential correlation $e^{-|t|/\tau}$ (whose area is $2\tau$) to white noise with covariance $2 \bm{\Sigma}\bm{\Sigma}^T \delta(t-s)$, yielding $2\bm{\Sigma}\bm{\Sigma}^T = 2\tau \bm{D}_0$ (see Appendix~\ref{sec:appendix_white_noise} for details).

To obtain a tractable reduced model for the observed slow variable, we replace the nonlocal (memory) drift in \eqref{eq:gle} with an effective Markovian drift $\vec F(\vec x)$. This Markovian closure absorbs the leading impact of the memory kernel into a state-dependent instantaneous drift. The function $\vec F$ and the noise amplitude matrix $\bm{\Sigma}$ are chosen so that the resulting stochastic model reproduces the empirical steady-state distribution and the short-lag time-correlation functions of $\vec x$.

For observations on time scales significantly longer than \(\tau\), we approximate the slow dynamics by an effective Langevin equation:
\begin{equation}
    \dot{\vec{x}} \approx \vec{F}(\vec{x}) + \sqrt{2}\bm{\Sigma}(\vec{x})\vec{\xi}(t),
    \label{eq:effective_dynamics}
\end{equation}
where \(\vec{\xi}(t) \in \mathbb{R}^{D_x}\) has independent standard components, \(\vec{F}(\vec{x}) \in \mathbb{R}^{D_x}\) is the effective drift, and \(\bm{\Sigma}(\vec{x}) \in \mathbb{R}^{D_x \times D_x}\) is the noise amplitude matrix.

\subsubsection{The Markovian Closure via Score-Based Modeling}

We defined \(\vec{F}\) and \(\bm{\Sigma}\) so that the reduced SDE \eqref{eq:effective_dynamics} reproduces both the observed probability density function (PDF) and the short-lag autocorrelation function (ACF) of $\vec x$ from data.

It has been shown that this projection-based model can be achieved with the following choices of drift and diffusion coefficients \cite{giorgini2025data}:
\begin{equation}
    \dot{\vec{x}} = \bm{\Phi} \nabla_{\vec{x}} \ln p_S(\vec{x}) + \sqrt{2}\,\text{chol}(\bm{\Phi}_S)\vec{\xi}(t),
    \label{eq:effective_dynamics_additive}
\end{equation}
where $p_S(\vec{x})$ is the stationary probability density, $\bm{\Phi}$ is a drift tensor, and $\bm{\Phi}_S = \frac{1}{2}(\bm{\Phi}+\bm{\Phi}^T)$ is its symmetric part. 

To derive this form, we start from a general autonomous It\^{o} SDE
\begin{equation}
    d\vec{x}(t) = \vec{F}(\vec{x}(t))dt + \sqrt{2}\bm{\Sigma}d\vec{W}(t),
    \label{eq:general_sde}
\end{equation}
where $\vec{F}: \mathbb{R}^{D_x} \to \mathbb{R}^{D_x}$ is the drift vector field and $\bm{\Sigma} \in \mathbb{R}^{D_x \times D_x}$ is a constant diffusion matrix. The time evolution of the probability density function $p(\vec{x}, t)$ is governed by the Fokker-Planck equation:
\begin{equation}
    \frac{\partial p}{\partial t} = -\sum_{i=1}^{D_x} \frac{\partial}{\partial x_i}[F_i(\vec{x})p] + \sum_{i,j=1}^{D_x} \frac{\partial^2}{\partial x_i \partial x_j}[(\bm{\Sigma}\bm{\Sigma}^{T})_{ij}p].
    \label{eq:fp_general}
\end{equation}
For an ergodic system admitting a unique stationary distribution $p_S(\vec{x})$, we impose $\partial_t p_S = 0$, yielding the stationary Fokker-Planck equation:
\begin{equation}
    -\vec{\nabla}\cdot[\vec{F}(\vec{x})p_{S}(\vec{x})] + \vec{\nabla}\cdot[\bm{\Sigma}\bm{\Sigma}^{T}\vec{\nabla}p_{S}(\vec{x})] = 0.
    \label{eq:fp_stationary}
\end{equation}
Using the identity $\vec{\nabla}p_S = p_S\vec{\nabla}\ln{p_S}$, we can rewrite Equation \eqref{eq:fp_stationary} as:
\begin{equation}
    \vec{\nabla}\cdot\left[\left(\vec{F}(\vec{x}) - \bm{\Sigma}\bm{\Sigma}^{T}\vec{\nabla}\ln{p_{S}(\vec{x})}\right)p_{S}(\vec{x})\right] = 0.
    \label{eq:fp_rearranged}
\end{equation}
This zero-divergence condition allows the drift to be decomposed as:
\begin{equation}
    \vec{F}(\vec{x}) = \bm{\Sigma}\bm{\Sigma}^{T}\vec{\nabla}\ln{p_{S}(\vec{x})} + \vec{g}(\vec{x}),
    \label{eq:drift_decomposition_general}
\end{equation}
where the vector field $\vec{g}(\vec{x})$ must satisfy $\vec{\nabla}\cdot(\vec{g}(\vec{x})p_{S}(\vec{x})) = 0$. The first term represents the conservative (reversible) drift aligned with the gradient of the log-probability, while $\vec{g}(\vec{x})$ captures non-conservative (irreversible) dynamics. A convenient parameterization is to model the irreversible part as $\vec{g}(\vec{x}) = \bm{R}(\vec{x})\vec{\nabla}\ln p_S(\vec{x})$ with antisymmetric $\bm{R}(\vec{x})$. We further approximate $\bm{R}(\vec{x})$ by a constant antisymmetric matrix $\bm{\Phi}_A$, effectively averaging over the state-dependent circulatory fluctuations. Defining the score function $\vec{s}(\vec{x}) \equiv \vec{\nabla}\ln p_S(\vec{x})$ and letting $\bm{\Phi} = \bm{\Sigma}\bm{\Sigma}^{T} + \bm{\Phi}_A$, we arrive at Equation \eqref{eq:effective_dynamics_additive}.

\subsubsection{Construction of the Score Function}
\label{subsubsec:score_function}
The score function, $\vec{s}(\vec{x}) = \nabla_{\vec{x}} \ln p_S(\vec{x})$, is the gradient of the logarithm of the stationary probability density $p_S(\vec{x})$. It is essential for defining the conservative part of the drift and ensuring the model reproduces the system's invariant measure. We estimate the score function using the K-means Gaussian Mixture Model (KGMM) method \cite{giorgini2025kgmm}.

The KGMM approach relies on the identity $\nabla_{\vec{x}} \ln p(\vec{x}) = -\frac{1}{\sigma_G} \mathbb{E}[\vec{z}|\vec{x}]$, where $\vec{x} = \vec{\mu} + \sigma_G \vec{z}$ with $\vec{z} \sim \mathcal{N}(\vec{0}, \bm{I})$, $\sigma_G$ is the Gaussian kernel width and $\vec{\mu}$ are the observed data points. The estimation involves:
\begin{enumerate}
    \item Perturbing data points $\{\vec{\mu}_i\}$ with Gaussian noise: $\vec{x}_i = \vec{\mu}_i + \sigma_G \vec{z}_i$.
    \item Partitioning the perturbed samples $\{\vec{x}_i\}$ into $N_C$ clusters $\{\Omega_j\}$ using bisecting K-means, with centroids $\vec{C}_j$.
    \item Estimating $\mathbb{E}[\vec{z}|\vec{x} \in \Omega_j]$ by averaging $\vec{z}_i$ for points in each cluster $\Omega_j$.
    \item Obtaining discrete score estimates $\vec{s}_j$ at centroids $\vec{C}_j$.
    \item Training a neural network to interpolate these discrete estimates for a continuous score function $\vec{s}(\vec{x})$.
\end{enumerate}
Further details on the KGMM method can be found in Giorgini et al. (2025) \cite{giorgini2025kgmm}.

\subsubsection{Construction of the Drift Tensor}
\label{subsubsec:drift_tensor}

The drift tensor $\bm{\Phi}$ in Equation \eqref{eq:effective_dynamics_additive} is constructed to reproduce the system's autocorrelation functions by capturing both conservative (reversible) and non-conservative (irreversible) dynamics. It is decomposed into a symmetric part, $\bm{\Phi}_S = \bm{\Sigma}\bm{\Sigma}^{T} = \frac{1}{2}(\bm{\Phi}+\bm{\Phi}^T)$, and an antisymmetric part, $\bm{\Phi}_A = \bm{\tilde{R}} = \frac{1}{2}(\bm{\Phi}-\bm{\Phi}^T)$, where $\bm{\tilde{R}}$ is an effective antisymmetric tensor representing averaged irreversible effects. The diffusion operator in Equation \eqref{eq:effective_dynamics_additive} is $\bm{\Phi}_S$, and the noise amplitude is $\sqrt{2}\,\text{chol}(\bm{\Phi}_S)$.

To construct $\bm{\Phi}$, we employ a finite-volume discretization of the state space using the same partition into $N_C$ control volumes $\{\Omega_j\}$ with centroids $\vec{C}_j$ as in the KGMM score estimation (Section \ref{subsubsec:score_function}). This ensures consistency between the score function and drift matrix estimation. We approximate the generator of the Perron–Frobenius operator with a rate matrix $\bm{Q} \in \mathbb{R}^{N_C \times N_C}$, which governs the evolution of the discrete probability vector $\vec{\rho}(t)$, where each component $\rho_j(t)$ represents the probability of the system being in control volume $\Omega_j$ at time $t$:
\begin{equation}
\dot{\vec{\rho}} = \bm{Q}\vec{\rho}.
\label{eq:rate_eq}
\end{equation}
The off-diagonal elements $Q_{ij}$ represent transition rates from volume $j$ to volume $i$, estimated from empirical transition counts in short-time trajectory data. The diagonal elements are determined by probability conservation:
\begin{equation}
Q_{ii} = -\sum_{j \neq i} Q_{ji}.
\end{equation}
The discrete stationary probability density function $\vec{\pi}$, with components $\pi_j$ representing the stationary probability mass in volume $\Omega_j$, satisfies $\bm{Q}\vec{\pi} = \vec{0}$.

Let $x_i^n$ denote the $i$-th component of the centroid $\vec{C}_n$. From the rate matrix $\bm{Q}$, the short-time correlation functions for small $\tau$ can be expanded as:
\begin{equation}
\lim_{\tau \to 0} C_{ij}(\tau)
=
\sum_{n=1}^{N_C}
x_j^n\,\pi_n
\sum_{m=1}^{N_C}
x_i^m \bigl[e^{\bm{Q}\,\tau}\bigr]_{mn}
\approx
\sum_{n=1}^{N_C}
x_j^n\,\pi_n
\sum_{m=1}^{N_C}
x_i^m \bigl[\bm{I} + \bm{Q}\,\tau\bigr]_{mn}.
\label{eq:discrete_acf}
\end{equation}
Simultaneously, the Langevin system \eqref{eq:effective_dynamics_additive} with drift $\bm{\Phi}\,\vec{s}(\vec{x})$ predicts:
\begin{equation}
\lim_{\tau \to 0} C_{ij}(\tau)
\approx
\sum_{n=1}^{N_C}
x_j^n\,\pi_n\Bigl( x_i^n
   + \bigl[\bm{\Phi}\,\vec{s}_n\bigr]_{i}\tau
\Bigr),
\label{eq:langevin_acf}
\end{equation}
where $\vec{s}_n = \nabla_{\vec{x}} \ln p_S(\vec{C}_n)$ is the score function evaluated at centroid $\vec{C}_n$, obtained from the KGMM method.

Matching the $\mathcal{O}(\tau)$ terms in Equations \eqref{eq:discrete_acf} and \eqref{eq:langevin_acf} yields the moment-matching condition:
\begin{equation}
\sum_{n,m=1}^{N_C}
x_j^n\,\pi_n \,x_i^m\,Q_{mn}
\;=\;
\sum_{n=1}^{N_C}
x_j^n\,\pi_n
\bigl[
\bm{\Phi}\,\vec{s}_n
\bigr]_i.
\end{equation}
Defining the moment matrices
\begin{equation}
M_{i,j}
:=
\sum_{n,m=1}^{N_C}
x_j^n\,\pi_n \,x_i^m\,Q_{mn},
\quad
V_{j,k}
:=
\sum_{n=1}^{N_C}
x_j^n\,\pi_n
\,[\vec{s}_n]_k,
\end{equation}
we obtain the matrix equation $\bm{M} = \bm{\Phi}\,\bm{V}^T$. Solving for $\bm{\Phi}$ using the Moore–Penrose pseudoinverse:
\begin{equation}
\bm{\Phi}
=
\bm{M}\,(\bm{V}^T)^{+}.
\label{eq:Phi_solution}
\end{equation}
This expression provides a systematic, data-driven construction of the drift matrix that preserves both the stationary distribution (through the score function) and short-time correlations (through the rate matrix $\bm{Q}$). For a detailed derivation, see Giorgini (2025) on data-driven decomposition of dynamics \cite{giorgini2025data}.

\subsubsection{Extension to Cyclo-Stationary Systems via State-Space Augmentation}
\label{subsubsec:cyclostationary_extension}

Many natural systems exhibit cyclo-stationary behavior, characterized by periodic forcing such as annual and diurnal cycles in climate systems, circadian rhythms in biological systems, or seasonal patterns in economic systems. Unlike the autonomous systems discussed above, these systems have dynamics that explicitly depend on time through periodic components. The framework presented can be extended to handle such cyclo-stationary systems through a state-space augmentation strategy.

The key idea is to recast the non-autonomous cyclo-stationary problem into an autonomous one in a higher-dimensional state space. Consider a system driven by $N$ known periodicities with angular frequencies $\{\omega_i\}_{i=1}^N$. We augment the $D_x$-dimensional physical state $\vec{x}_{\mathrm{phys}}(t)$ with $2N$ harmonic coordinates:
\begin{equation}
\vec{x}_{\mathrm{aug}}(t)
=
\big(\sin(\omega_1 t),\,\cos(\omega_1 t),\,\dots,\,\sin(\omega_N t),\,\cos(\omega_N t),\,\vec{x}_{\mathrm{phys}}(t)^\top\big)^\top
\in \mathbb{R}^{2N+D_x}.
\label{eq:augmented_state}
\end{equation}
These additional coordinates, often called "clock variables," explicitly encode the phase of each periodic forcing. In the augmented space, we can construct a single, constant drift matrix $\bm{\Phi}_{\mathrm{aug}}$ for the entire system, which implicitly encodes the time-dependent dynamics of the original physical variables while preserving the conservative-non-conservative decomposition.

The methodology proceeds analogously to the autonomous case: we estimate the score function $\vec{s}_{\mathrm{aug}}(\vec{x}_{\mathrm{aug}})$ on the augmented state using KGMM and construct the drift matrix $\bm{\Phi}_{\mathrm{aug}}$ using moment matching. The symmetric part $\bm{S}_{\mathrm{aug}} = \frac{1}{2}(\bm{\Phi}_{\mathrm{aug}}+\bm{\Phi}_{\mathrm{aug}}^{\!\top})$ determines the diffusion matrix through $\bm{\Sigma}_{\mathrm{aug}}\bm{\Sigma}_{\mathrm{aug}}^{\!\top} = \bm{S}_{\mathrm{aug}}$.

After learning the augmented model, we integrate only the $D_x$ physical coordinates while evaluating the full augmented score at the current clock phase. Defining the time-dependent score on the physical state by
\begin{equation}
\vec{s}\!\left(\vec{x}_{\mathrm{phys}}(t),t\right)
=
\vec{s}_{\mathrm{aug}}\!\left(\vec{c}(t),\,\vec{x}_{\mathrm{phys}}(t)\right),
\qquad
\vec{c}(t)
=
\big(\sin(\omega_1 t),\cos(\omega_1 t),\ldots,\sin(\omega_N t),\cos(\omega_N t)\big),
\label{eq:time_dependent_score}
\end{equation}
and extracting the physical block-rows of the operators, the SDE for the physical coordinates becomes
\begin{equation}
\mathrm{d}\vec{x}_{\mathrm{phys}}(t)
=
\bm{\Phi}_{\mathrm{phys}}\,
\vec{s}\!\left(\vec{x}_{\mathrm{phys}}(t),t\right)\,\mathrm{d}t
+
\sqrt{2}\,\bm{\Sigma}_{\mathrm{phys}}\,\mathrm{d}\vec{W}_t,
\label{eq:phys_sde_cyclostat}
\end{equation}
where $\bm{\Phi}_{\mathrm{phys}}$ and $\bm{\Sigma}_{\mathrm{phys}}$ correspond to the rows associated with the physical coordinates in the augmented operators. By pinning the clock to the true phase $\vec{c}(t)$ and evolving only the physical coordinates with the full augmented score evaluated on that slice, the resulting process is cyclo-stationary with the same fundamental periodicities as the data. This approach bypasses the need for time-dependent parameters by embedding the temporal dependence into the structure of the joint probability distribution in the augmented space, while preserving both the invariant measure and the physically meaningful decomposition of conservative and non-conservative dynamics.

\subsection{Dynamics of the Fast Variable}
\label{subsec:fast_variable_dynamics}

We extract a normalized fast signal $\widetilde{\vec{\xi}}(t)$ directly from the observed slow variable $\vec x(t)$ and the learned score closure, apply variance normalization to obtain $\widetilde{\vec{\xi}}'(t)$, and then learn a continuous-time surrogate for the variance-normalized signal via delay embedding. The short-horizon reduced dynamics define how $\widetilde{\vec{\xi}}(t)$ drives $\vec x$:
\begin{equation}
    \dot{\vec{x}}(t) = \bm{\Phi} \, \vec s\big(\vec{x}(t)\big) + \frac{1}{\sqrt{\tau}}\,\text{chol}\!\big(\bm{\Phi}_S\big)\,\widetilde{\vec{\xi}}(t),
    \label{eq:effective_dynamics_short}
\end{equation}
where $\vec s(\vec x)=\nabla_{\vec x}\ln p_S(\vec x)$, $\bm{\Phi}_S = \frac{1}{2}(\bm{\Phi}+\bm{\Phi}^T)$, and $\tau$ is the characteristic decorrelation time. Writing the residual $\vec r(t) := \dot{\vec x}(t) - \bm{\Phi}\,\vec s\big(\vec x(t)\big)$ and the gain matrix $\mathbf{C} := (1/\sqrt{\tau})\,\text{chol}\!\big(\bm{\Phi}_S\big)$, the fast signal follows from the algebraic inversion
\begin{equation}
    \widetilde{\vec{\xi}}(t) = \mathbf{C}^{-1}\,\vec r(t) \;=\; \sqrt{\tau}\,\text{chol}\!\big(\bm{\Phi}_S\big)^{-1}\,\big(\dot{\vec x}(t) - \bm{\Phi}\,\vec s(\vec x(t))\big),
    \label{eq:fast_signal_inversion}
\end{equation}
which normalizes $\widetilde{\vec{\xi}}$ to time-integrated covariance $2\tau \mathbf{I}$ (equivalently, $\widetilde{\vec{\xi}} = \sqrt{2\tau}\,\vec{\xi}$ under the closure). If $\bm{\Phi}_S$ is ill-conditioned, we use the regularized factor $\text{chol}\!\big(\bm{\Phi}_S+\varepsilon\mathbf{I}\big)^{-1}$ with small $\varepsilon>0$. In discrete time with uniform step $\Delta t$, we estimate $\dot{\vec x}$ using robust differentiators (e.g., Savitzky--Golay or Tikhonov collocation), evaluate the score with the KGMM network, form $\vec r$, and compute $\widetilde{\vec{\xi}}=\mathbf{C}^{-1}\vec r$. The sampling interval should resolve the slow timescale yet not be so small that derivative estimates are noise-dominated (a pragmatic choice is near the knee of the $\dot{\vec x}$ power spectrum). We obtain the scalar $\tau$ by evaluating the integrated autocorrelation time of the slow variable $\vec x(t)$.

Although equation~\eqref{eq:fast_signal_inversion} ensures that $\widetilde{\vec{\xi}}$ has time-integrated covariance $2\tau \mathbf{I}$ when averaged over the invariant measure of $\vec{x}$, the fast signal exhibits significant state dependence: both its instantaneous amplitude and local statistical properties depend on the slow variable $\vec{x}(t)$. This dependence arises because the fast dynamics are conditioned on the current position along the slow manifold, with the local geometry and stability of the slow flow modulating the characteristics of the fast fluctuations. On sufficiently long timescales—much longer than the decorrelation time $\tau$—the state-averaged statistics dominate and the $\vec{x}$-dependence can be neglected. However, for short-horizon forecasts where we aim to predict transient dynamics and rare transitions, accurately capturing how $\widetilde{\vec{\xi}}(t)$ evolves in response to the current slow state $\vec{x}(t)$ is essential. This state dependence motivates conditioning the fast-dynamics surrogate explicitly on $\vec{x}(t)$.

A practical complication is that the conditional variance $\mathrm{Var}\big(\widetilde{\vec{\xi}} \mid \vec{x}\big)$ can vary substantially across the state space, even though the global average variance is unity by construction. Such heteroskedasticity presents challenges for neural network training: regions of high variance disproportionately influence the loss function, gradient magnitudes fluctuate widely, and the network may underfit quieter regions of state space. To address this, we decompose the fast signal into a state-dependent amplitude modulation and a homoskedastic residual:
\begin{equation}
    \widetilde{\vec{\xi}}(t) = g\big(\vec{x}(t)\big) \, \widetilde{\vec{\xi}}'(t),
    \label{eq:variance_normalization}
\end{equation}
where $g(\vec{x}): \mathbb{R}^{D_x} \to \mathbb{R}_{>0}$ is a scalar gain function ensuring that $\mathrm{Var}\big(\widetilde{\vec{\xi}}' \mid \vec{x}\big) = 1$ for all $\vec{x}$, and $\widetilde{\vec{\xi}}'(t)$ is the variance-normalized fast signal with constant unit variance at every point in state space. In practice, we estimate $g(\vec{x})$ from training data by computing the empirical conditional standard deviation via binning, kernel smoothing, or by training a small auxiliary neural network to predict $\sqrt{\mathbb{E}[\widetilde{\vec{\xi}}^2 \mid \vec{x}]}$. Training the fast-dynamics surrogate on the homoskedastic signal $\widetilde{\vec{\xi}}'(t)$ yields more stable optimization, more uniform convergence across the state space, and improved generalization. Once the surrogate for $\widetilde{\vec{\xi}}'(t)$ is learned, we reconstruct the original fast signal via $\widetilde{\vec{\xi}}(t) = g\big(\vec{x}(t)\big) \widetilde{\vec{\xi}}'(t)$ during inference.

To represent the fast dynamics, we construct delay vectors $\mathbf{z}'(t) = [\widetilde{\vec{\xi}}'(t),\,\widetilde{\vec{\xi}}'(t-\Delta\tau),\,\ldots,\,\widetilde{\vec{\xi}}'(t-(m-1)\Delta\tau)] \in \mathbb{R}^{m\times D_x}$ from the variance-normalized signal and learn a Neural ODE $\frac{d}{dt}{\widetilde{\vec{\xi}}}'(t)=\vec G_\theta\big(\mathbf{z}'(t),\vec{x}(t)\big)$ that explicitly conditions on the slow variable $\vec{x}(t)$. Training minimizes a multistep prediction loss together with statistical penalties that enforce $\mathbb{E}[\widetilde{\vec{\xi}}']=\vec 0$ and $\mathrm{Cov}(\widetilde{\vec{\xi}}')=\mathbf{I}$ on long free runs:
\begin{equation}
    \mathcal{L}_{\text{pred}}(\bm{\theta}) = \frac{1}{N} \sum_{i=1}^{N} \sum_{j=1}^{T} \big\| \widetilde{\vec{\xi}}'_{i+j} - \widetilde{\vec{\xi}}'^{\,\text{pred}}_{i+j}(\bm{\theta}) \big\|^2.
    \label{eq:NODE_loss}
\end{equation}
The delay-embedding methodology is effective when the fast driver is predominantly deterministic and smooth. According to Takens' delay embedding theorem \cite{Takens}, the embedding dimension $m$ and delay $\Delta\tau$ are chosen such that $m \geq 2d_A + 1$, where $d_A$ is the correlation dimension of the attractor, ensuring that the delay-coordinate map provides a diffeomorphic reconstruction of the underlying attractor. In practice, $\Delta\tau$ is selected to avoid redundancy or decorrelation, for instance by choosing $\Delta\tau$ at the first minimum of average mutual information or near the $e^{-1}$ decorrelation time. Approximate stationarity of $\widetilde{\vec{\xi}}'(t)$ over the training window and uniform sampling are necessary; significant measurement noise may require larger $m$ and regularization in $\vec G_\theta$. Noninvertible gains in the normalization are alleviated by the aforementioned $\varepsilon$-regularization. If known periodic or cyclostationary forcings are present, augmenting $\mathbf{z}'(t)$ with $\sin(\omega_i t)$ and $\cos(\omega_i t)$ restores effective autonomy. When $\widetilde{\vec{\xi}}'$ contains substantial truly random forcing or the fast attractor is poorly sampled, delay maps approximate correlations rather than a diffeomorphic copy; this is detectable via degraded one-step and free-run statistics and unstable correlation-dimension estimates under data thinning.

For short-horizon forecasts, we integrate
\begin{equation}
\label{eq:effective_dynamics_NODE}
  \dot{\vec x}(t)=\bm{\Phi}\,\vec s\big(\vec x(t)\big)+\frac{1}{\sqrt{\tau}}\,\text{chol}\!\big(\bm{\Phi}_S\big)\,\widehat{\vec{\xi}}(t),
\end{equation}
where $\widehat{\vec{\xi}}(t) = g\big(\vec{x}(t)\big) \widehat{\vec{\xi}}'(t)$ is reconstructed from the variance-normalized signal $\widehat{\vec{\xi}}'(t)$ generated by the trained NODE from the delay-embedded state. We validate the NODE surrogate by verifying that $\widehat{\vec{\xi}}'(t)$ accurately reproduces the observed trajectories of $\widetilde{\vec{\xi}}'(t)$ up to one Lyapunov time, and that the rescaled signal $\widehat{\vec{\xi}}(t)$ matches the original $\widetilde{\vec{\xi}}(t)$. The Lyapunov time, defined as the inverse of the largest Lyapunov exponent, characterizes the timescale over which nearby trajectories in a chaotic system diverge exponentially; it represents the intrinsic predictability horizon beyond which deterministic forecasts become unreliable regardless of model fidelity. Additionally, we verify that the NODE-generated time series reproduces both the stationary probability density function (PDF) and the autocorrelation function (ACF) of the observed chaotic signal $\widetilde{\vec{\xi}}(t)$, ensuring that the surrogate captures not only short-term dynamical accuracy but also the correct long-term statistical properties of the fast forcing. With the present state-independent diffusion factor, It\^{o} and Stratonovich interpretations coincide and no noise-induced drift arises.

We apply this hybrid framework to predict rare transitions in metastable systems. Metastable systems are characterized by multiple quasi-stable states separated by energy barriers, in which the dynamics spend long periods near local minima of an effective potential before undergoing rapid, noise-driven transitions between basins of attraction. As discussed in the Introduction, such transitions are critical events in domains ranging from climate tipping points to phase changes in materials, and their prediction requires resolving the fast transients that trigger escape from metastable wells—a task for which Gaussian white-noise approximations often prove insufficient. By coupling the NODE-based chaotic forcing with the score-informed closure, we perform short-horizon forecasts initialized at progressively earlier times before observed critical transitions, thereby assessing the model's ability to anticipate these rare but consequential events.

\subsubsection{Extension to Non-Autonomous Chaotic Forcing.}

If the fast chaotic process $\widetilde{\vec{\xi}}'(t)$ exhibits explicit time dependence due to periodic or cyclo-stationary forcings (e.g., seasonal modulation of atmospheric turbulence), the NODE must be modified to capture this non-autonomous behavior. Following the state-space augmentation strategy described in Section~\ref{subsubsec:cyclostationary_extension}, we augment the delay-embedded state with harmonic clock variables corresponding to the known periodicities $\{\omega_i\}_{i=1}^N$. The augmented delay vector becomes
\begin{equation}
\mathbf{z}'_{\mathrm{aug}}(t) = \big[\,\vec{c}(t),\,\mathbf{z}'(t)\,\big] \in \mathbb{R}^{2N+m\times D_x},
\label{eq:augmented_delay_vector}
\end{equation}
where $\vec{c}(t) = \big(\sin(\omega_1 t),\,\cos(\omega_1 t),\,\ldots,\,\sin(\omega_N t),\,\cos(\omega_N t)\big)$ is the clock-variable vector and $\mathbf{z}'(t) = \big[\widetilde{\vec{\xi}}'(t),\,\widetilde{\vec{\xi}}'(t-\Delta\tau),\,\ldots,\,\widetilde{\vec{\xi}}'(t-(m-1)\Delta\tau)\big]$ is the delay-embedded variance-normalized fast-signal vector.
The Neural ODE is trained to predict $\frac{d}{dt}{\widetilde{\vec{\xi}}}'(t)=\vec G_\theta\big(\mathbf{z}'_{\mathrm{aug}}(t),\vec{x}(t)\big)$ with the augmented input. This augmentation recasts the non-autonomous problem as an autonomous one in the higher-dimensional delay space, allowing the NODE to learn how the fast dynamics vary with the phase of the external forcing. Alternatively, one can directly include time as an explicit input $\frac{d}{dt}{\widetilde{\vec{\xi}}}'(t)=\vec G_\theta\big(\mathbf{z}'(t),\vec{x}(t),t\big)$ and train the network to recognize periodic patterns, though this approach may require careful regularization to avoid overfitting to the specific time window of the training data. The augmented approach is generally more robust as it explicitly encodes the periodicity structure, ensuring that the learned NODE respects the known cyclo-stationary character of the forcing and generalizes correctly to arbitrary time horizons. During inference, the clock variables are evaluated at the current simulation time, providing the NODE with the appropriate seasonal or diurnal phase information needed to generate realistic fast forcing trajectories that reflect the time-varying statistical properties of the chaotic driver.

\section{Numerical Experiments}
\label{sec:experiments}

We validate the hybrid modeling framework developed in Section~\ref{sec:methods} on four prototypical metastable dynamical systems of increasing complexity. These systems encompass bistable and tristable potentials subject to chaotic fast forcing with both additive and multiplicative coupling, as well as autonomous and non-autonomous configurations with cycloperiodic drift components. This hierarchy of test cases allows systematic assessment of the method's ability to reproduce both long-term statistical fidelity and short-term predictive accuracy, including the anticipation of rare critical transitions between metastable states.

All four systems share a common mathematical structure: the dynamics of the observed slow variable $x(t) \in \mathbb{R}$ are coupled to a three-dimensional fast chaotic process $\vec{y}(t) = (y_1(t), y_2(t), y_3(t))$ governed by the Lorenz 63 equations. The general form of the coupled system is
\begin{equation} \label{eq:general_coupled_system}
\left\{
\begin{aligned}
\dot{x}(t) &= f\!\big(x(t),t\big) + g\!\big(x(t)\big)\,y_2(t), \\
\dot{y}_1(t) &= \frac{10}{\varepsilon^2} \bigl[ y_2(t) - y_1(t) \bigr], \\
\dot{y}_2(t) &= \frac{1}{\varepsilon^2} \bigl[ 28\,y_1(t) - y_2(t) - y_1(t)\,y_3(t) \bigr], \\
\dot{y}_3(t) &= \frac{1}{\varepsilon^2} \bigl[ y_1(t)\,y_2(t) - \tfrac{8}{3}\,y_3(t) \bigr],
\end{aligned}
\right.
\end{equation}
where $f(x,t)$ represents the deterministic drift of the slow variable (which may include autonomous potential-derived terms and non-autonomous periodic forcing), and $g(x)$ is the state-dependent amplification function modulating the influence of the fast chaotic signal $y_2(t)$ on the slow dynamics. The fast subsystem evolves on the classical Lorenz 63 attractor with standard parameters (Prandtl number 10, Rayleigh number 28, aspect ratio $8/3$). The dimensionless parameter $\varepsilon$ controls the timescale separation: the fast variables evolve on a characteristic timescale $\mathcal{O}(\varepsilon^2)$, while the slow variable $x$ evolves on a timescale $\mathcal{O}(1)$. For $\varepsilon \ll 1$, the fast subsystem rapidly decorrelates relative to the slow dynamics, justifying the statistical closure approximations introduced in Section~\ref{subsec:slow_variable_dynamics}. Throughout all experiments we fix $\varepsilon = 0.5$, yielding a timescale separation sufficient for the colored-noise approximation to hold while maintaining computational tractability. All systems are integrated using a uniform time step $\Delta t = 0.001$ with initial conditions $x(0) = 0$ and $y_i(0)$ drawn uniformly from $[-10, 10]$ for $i = 1, 2, 3$.

Our validation protocol proceeds in two stages for each system, following the methodology detailed in Section~\ref{sec:methods}. First, we construct the long-term statistical closure by estimating the score function $\vec{s}(x) = \nabla_x \ln p_S(x)$ via the KGMM method and computing the drift tensor $\bm{\Phi}$ through moment matching on the finite-volume discretization. This yields an effective Langevin equation of the form $\dot{x} = \bm{\Phi}\,\vec{s}(x) + \sqrt{2}\,\text{chol}(\bm{\Phi}_S)\,\xi(t)$, where $\xi(t)$ is standard Gaussian white noise. We validate this reduced-order model (ROM) by integrating long trajectories and comparing the resulting steady-state probability density function (PDF) and autocorrelation function (ACF) against empirical statistics from the full system. Where analytical expressions for the score function are available, we use them as additional benchmarks. For the first system only, an analytical expression for the invariant measure is also available under the Gaussian white-noise approximation, providing a rigorous benchmark for assessing the data-driven closure. Second, we replace the Gaussian white-noise approximation with an explicit surrogate for the fast chaotic forcing by extracting the normalized fast signal $\widetilde{\vec{\xi}}(t)$ from the observed slow-variable time series via Equation~\eqref{eq:fast_signal_inversion}, applying variance normalization to obtain the homoskedastic signal $\widetilde{\vec{\xi}}'(t)$ as described in Section~\ref{subsec:fast_variable_dynamics}, constructing delay-embedded vectors, and training a Neural ODE to learn the vector field $\frac{d}{dt}{\widetilde{\vec{\xi}}}'(t) = \vec{G}_\theta(\mathbf{z}'(t))$. In the prototypical systems examined here, the fast signal depends on the slow variable only through its amplitude modulation, which is explicitly captured by the variance-normalization procedure; for computational simplicity, we therefore omit the slow variable $x(t)$ as a direct input to the NODE. As detailed in Section~\ref{subsec:fast_variable_dynamics}, this framework can be readily generalized to systems exhibiting more complex state-dependent fast dynamics by including $x(t)$ as an additional conditioning variable in the NODE architecture. For systems with periodic forcing of angular frequency $\omega$, the treatment depends on the relative timescale of the periodic component. When the forcing period is comparable to the fast-process decorrelation time (high-frequency regime), the periodic component is incorporated into the fast forcing and modeled by the NODE; in this case, the delay-embedded vectors are augmented with harmonic clock variables $(\sin(\omega t), \cos(\omega t))$ as described in Section~\ref{subsec:fast_variable_dynamics}, enabling the NODE to capture the time-dependent modulation of the chaotic signal. Conversely, when the forcing period is much longer than the fast decorrelation time (low-frequency regime), the periodic component couples to the slow dynamics and is incorporated into the statistical closure; in this case, the physical state $x(t)$ is augmented with the harmonic clock variables to form an extended state vector, and the KGMM score estimation is performed on these augmented coordinates as detailed in Section~\ref{subsubsec:cyclostationary_extension}. The trained NODE is validated by verifying that it reproduces both the statistical properties (PDF and ACF) and short-term trajectories of the observed fast signal up to approximately one Lyapunov time for the Lorenz 63 subsystem. Finally, we assess the short-horizon predictive capability of the hybrid model by integrating $\dot{x}(t) = \bm{\Phi}\,\vec{s}(x(t)) + (1/\sqrt{\tau})\,\text{chol}(\bm{\Phi}_S)\,\widehat{\vec{\xi}}(t)$, where $\widehat{\vec{\xi}}(t) = g(x(t))\widehat{\vec{\xi}}'(t)$ is the reconstructed fast signal generated by the NODE. For each critical transition observed in the validation dataset, we initialize forecasts at multiple lead times ranging from $0.25$ to $5$ time units before the transition and quantify predictive skill using root mean square error (RMSE) and temporal correlation between the ensemble-mean forecast (averaged over five to ten independently trained NODE instances) and the ground truth. This ensemble approach accounts for the intrinsic unpredictability beyond the Lyapunov time and provides uncertainty estimates for the forecasts.

\subsection{Bistable System with Additive Chaotic Forcing}
\label{subsec:double_well_lorenz}

The first system consists of a slow variable evolving in a symmetric double-well potential with additive chaotic forcing from the Lorenz 63 attractor. The slow-variable dynamics are governed by
\begin{equation} \label{eq:lorenz63}
\dot{x}(t) = x(t)\bigl[1 - x^2(t)\bigr] + \frac{\sigma}{\varepsilon} y_2(t),
\end{equation}
where the deterministic drift $f(x) = x(1 - x^2) = -U'(x)$ derives from the double-well potential $U(x) = -\frac{1}{2}x^2 + \frac{1}{4}x^4$ with symmetric minima at $x = \pm 1$, and the coupling function $g(x) = \sigma/\varepsilon$ is constant, yielding additive forcing. The fast subsystem follows Equation~\eqref{eq:general_coupled_system} with $\varepsilon = 0.5$ and coupling strength $\sigma = 0.08$. Under the timescale-separation approximation, the effective dynamics reduce to a Langevin equation $\dot{x}(t) = -U'(x) + \sqrt{2D_{\text{eff}}}\,\xi(t)$, where $D_{\text{eff}} = (\sigma^2/\varepsilon^2) \text{Var}[y_2] \cdot \tau_{\text{ACF}}$ is the effective diffusion coefficient (computed as $\text{Var}[y_2]$ times the integral of the normalized ACF of $y_2(t)$), and $\xi(t)$ is standard Gaussian white noise. From the stationary Fokker--Planck equation, the analytical score function is
\begin{equation}
\vec{s}(x) = \frac{d}{dx} \ln p_S(x) = \frac{-U'(x)}{D_{\text{eff}}} = \frac{\varepsilon^2}{\sigma^2 \text{Var}[y_2] \tau_{\text{ACF}}}\, x\bigl(1 - x^2\bigr),
\end{equation}
and the corresponding steady-state probability density is
\begin{equation}
\label{eq:realPDF}
p_S(x) = \frac{1}{Z} \exp\left(\!-\frac{U(x)}{D_{\text{eff}}}\right) = \frac{1}{Z} \exp\left(\frac{\varepsilon^2}{\sigma^2 \text{Var}[y_2] \tau_{\text{ACF}}}\left(\frac{x^2}{2} - \frac{x^4}{4}\right)\right),
\end{equation}
where $Z = \int_{-\infty}^{\infty} \exp\big(\varepsilon^2[\sigma^2 \text{Var}[y_2] \tau_{\text{ACF}}]^{-1}(x^2/2 - x^4/4)\big)\,dx$ is the partition function. These analytical expressions provide rigorous benchmarks for assessing the accuracy of the data-driven score and invariant measure obtained via KGMM. It should be noted that Equation~\eqref{eq:realPDF} assumes strictly Gaussian white noise forcing; since $y_2(t)$ exhibits mild non-Gaussian features and colored-noise correlations, small deviations between the analytical and observed distributions are expected.

\begin{figure}[H]
    \centering
    \includegraphics[width=\textwidth]{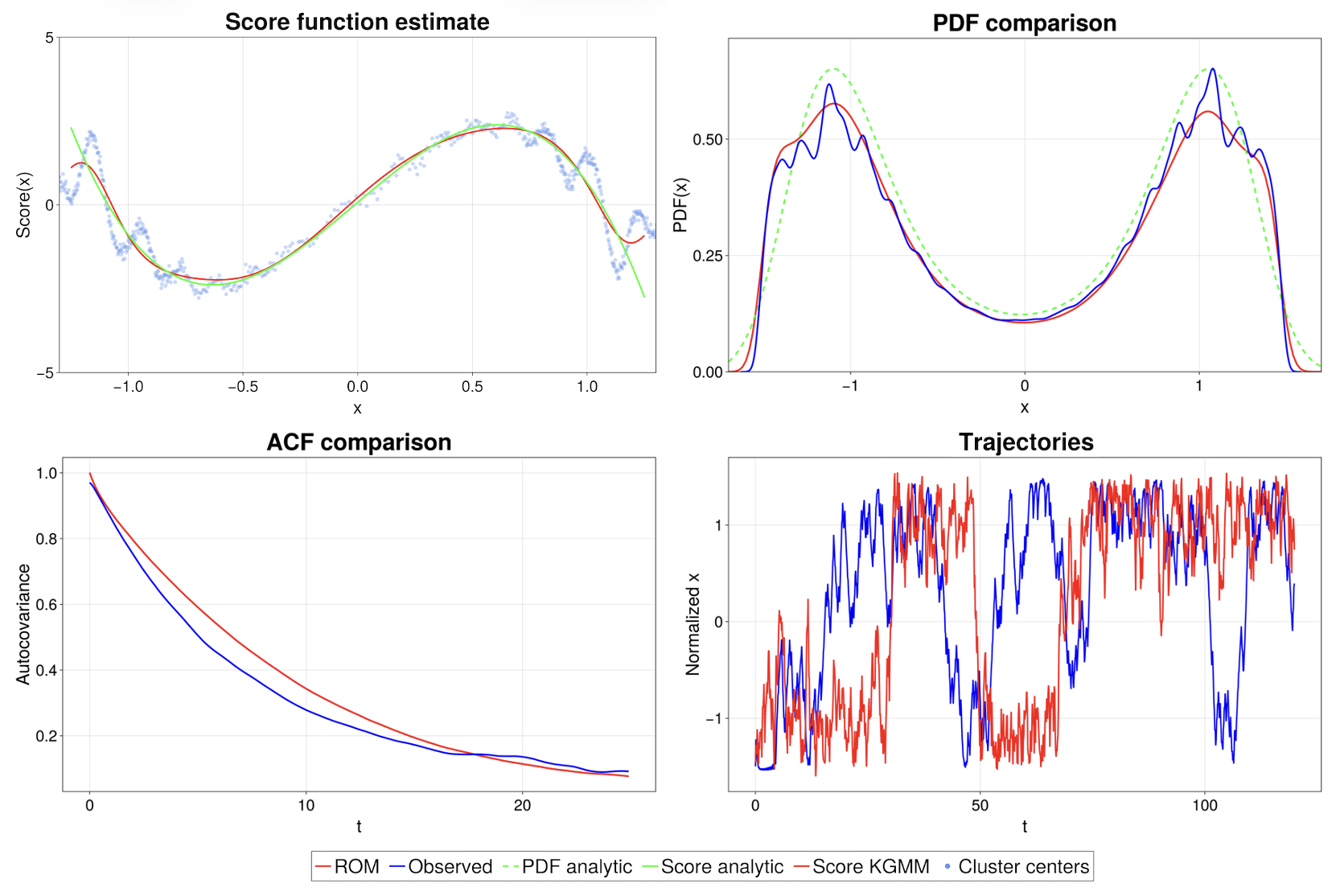}
\caption{Reduced-order model with Gaussian noise for system \ref{eq:lorenz63}. \textbf{Top row:} comparison between the score function estimated by the KGMM network (red) and the analytical expression $s(x)$ (green) for $\sigma_G = 0.05$, together with the stationary PDF showing three curves: the ROM reconstruction from the estimated score (red), the empirical distribution from observations of the full system (blue), and the analytical invariant measure assuming Gaussian white noise forcing (green). \textbf{Bottom row:} comparison between the ACF of the trajectory obtained by integrating the effective Langevin dynamics (red) and that of the observed trajectory (blue), and between trajectories generated by the effective Langevin model (red) and the observed trajectory from the original system defined by Eq.~\ref{eq:lorenz63} (blue) over an extended time horizon.}
\label{fig:reduced_model_summary}
\end{figure}

We estimate the score function using KGMM with Gaussian perturbation scale $\sigma_G = 0.05$ and construct the drift tensor $\bm{\Phi}$ via moment matching. The top panels of Figure~\ref{fig:reduced_model_summary} demonstrate excellent agreement between the KGMM-estimated score (red) and the analytical expression (green). The right panel shows three curves for the steady-state PDF: the analytical invariant measure (green), the empirical distribution from observations of the full system (blue), and the ROM reconstruction (red). Importantly, the analytical PDF (green) is derived under the assumption of Gaussian white noise forcing for the fast process; therefore, it exhibits slight deviations from the observed distribution, which is computed using the correct chaotic Lorenz 63 process as fast forcing. The data-driven ROM accurately captures the true empirical statistics. The bottom panels show that the ROM accurately reproduces both the autocorrelation function and representative long-horizon trajectories of $x(t)$, confirming statistical consistency over extended time horizons.

\begin{figure}[H]
\centering
    \includegraphics[width=1\textwidth]{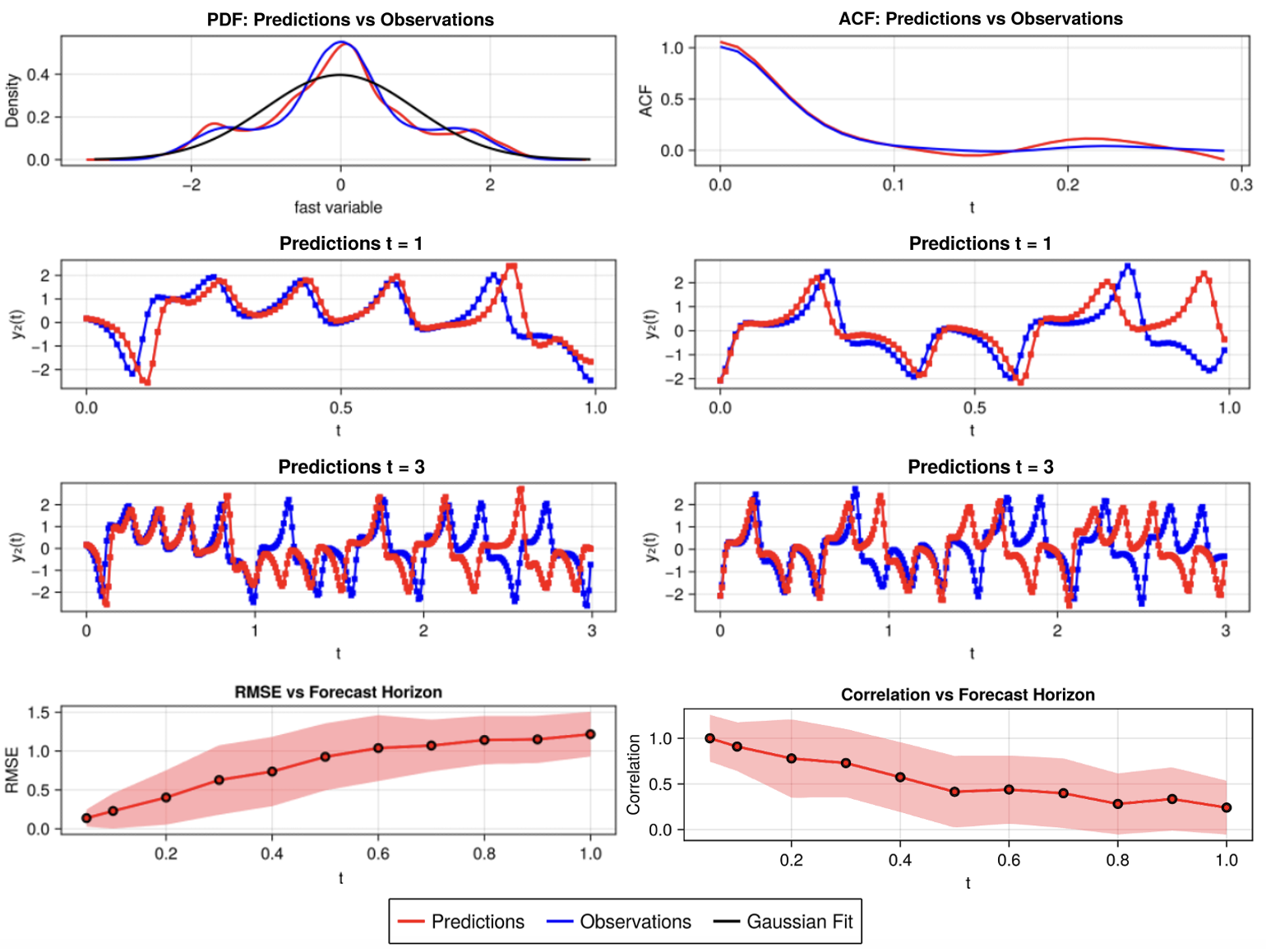}
    \caption{Performance of the NODE for system \ref{eq:lorenz63}. \textbf{Upper row:} statistical comparison between the trajectory generated by the trained NODE and the ground-truth trajectory obtained by integrating the Lorenz 63 subsystem; the panels show the probability density function (left) and autocorrelation function (right). \textbf{Central rows:} trajectories predicted by the NODE (red) vs.\ ground truth (blue) from different initial conditions, with short and intermediate predictive horizons; deviations grow beyond the Lyapunov time. \textbf{Bottom row:} RMSE versus predictive horizon, averaged over 200 trajectories with the shaded band indicating one standard deviation across trajectories.}
\label{fig:NODE_forecastLorenz63}
\end{figure}

For short-term forecasting, we train a Neural ODE on delay-embedded vectors of the extracted fast signal $\widetilde{\vec{\xi}}'(t)$ with embedding dimension $m = 10$, delay $\Delta\tau = 0.018$, and training minibatch duration $T = 0.6$. Figure~\ref{fig:NODE_forecastLorenz63} validates the NODE surrogate: the generated trajectories reproduce the PDF and ACF of $y_2(t)$ with high fidelity, and short-term trajectory predictions remain accurate up to approximately $t = 0.6$, approaching the Lyapunov time, beyond which RMSE grows and correlation decays as expected for chaotic dynamics.

\begin{figure}[H]
    \centering
  \includegraphics[width=0.7\linewidth]{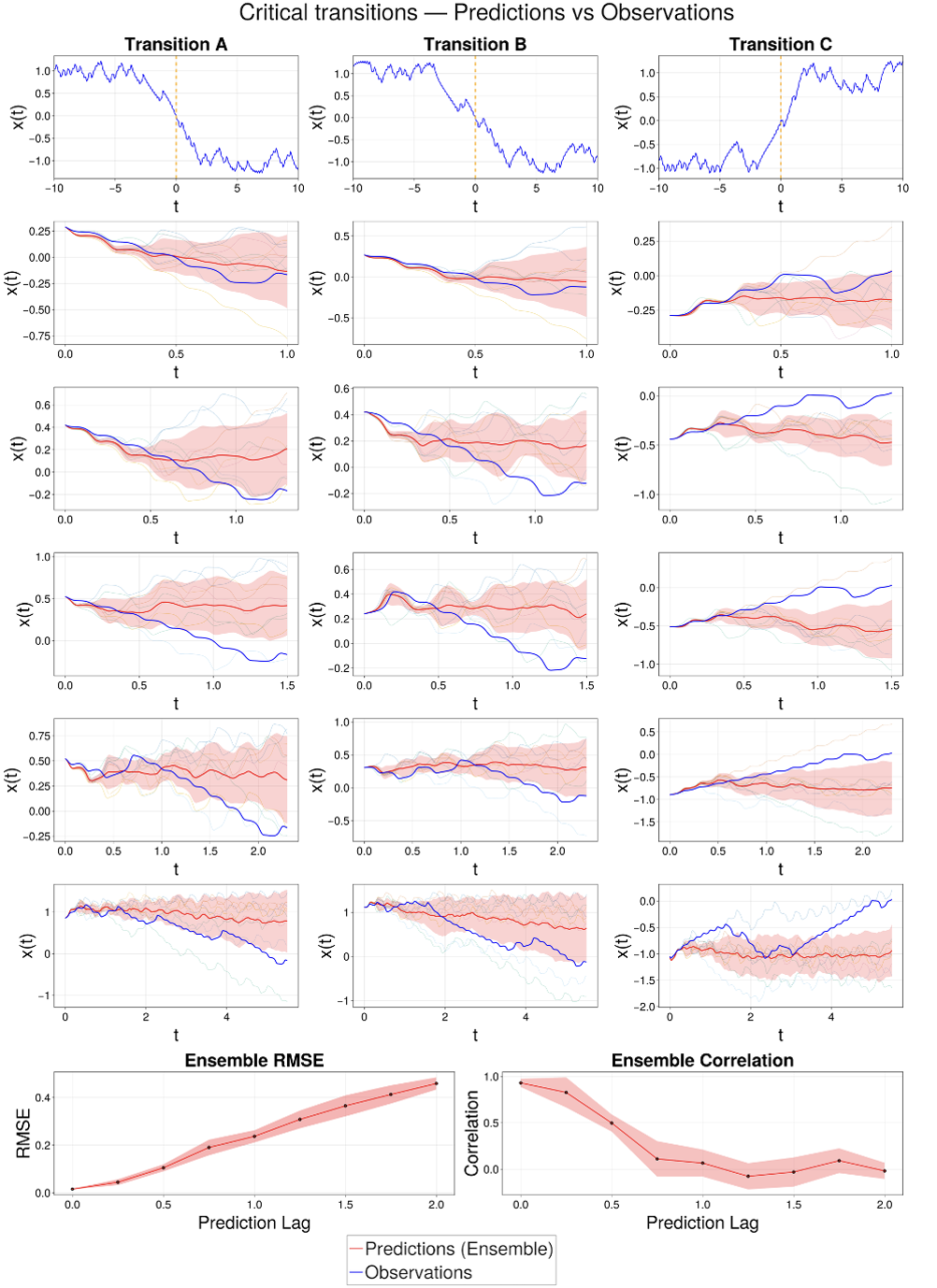}
\caption{\textbf{Top row:} three representative transitions with the transition time highlighted (dashed line); \textbf{Middle row:} model-ensemble forecasts for three transitions and five lags (left to right); \textbf{Bottom Row:} ensemble RMSE as a function of the forecast horizon.}
    \label{fig:full_forecast_overview}
\end{figure}

We assess the hybrid model's ability to anticipate critical transitions between the two metastable wells by initializing ensemble forecasts (averaged over ten independent NODE instances) at lead times $\{0.25, 0.5, 0.75, 1.0, 1.25, 1.5, 1.75, 2.0, 5.0\}$ before observed transitions in the validation dataset. Figure~\ref{fig:full_forecast_overview} presents three representative transitions: the top row shows ground-truth trajectories with transition times marked, the middle rows display ensemble forecasts initialized at multiple lags, and the bottom panel quantifies predictive skill via ensemble-mean RMSE as a function of forecast horizon. The results demonstrate that the hybrid framework successfully anticipates transitions up to approximately $t = 0.5$ in advance; beyond this lead time, an increasing fraction of ensemble members diverge from the ground truth, reflected in rising RMSE and declining correlation.

\begin{figure}[H]
    \centering
    \includegraphics[width=1\textwidth]{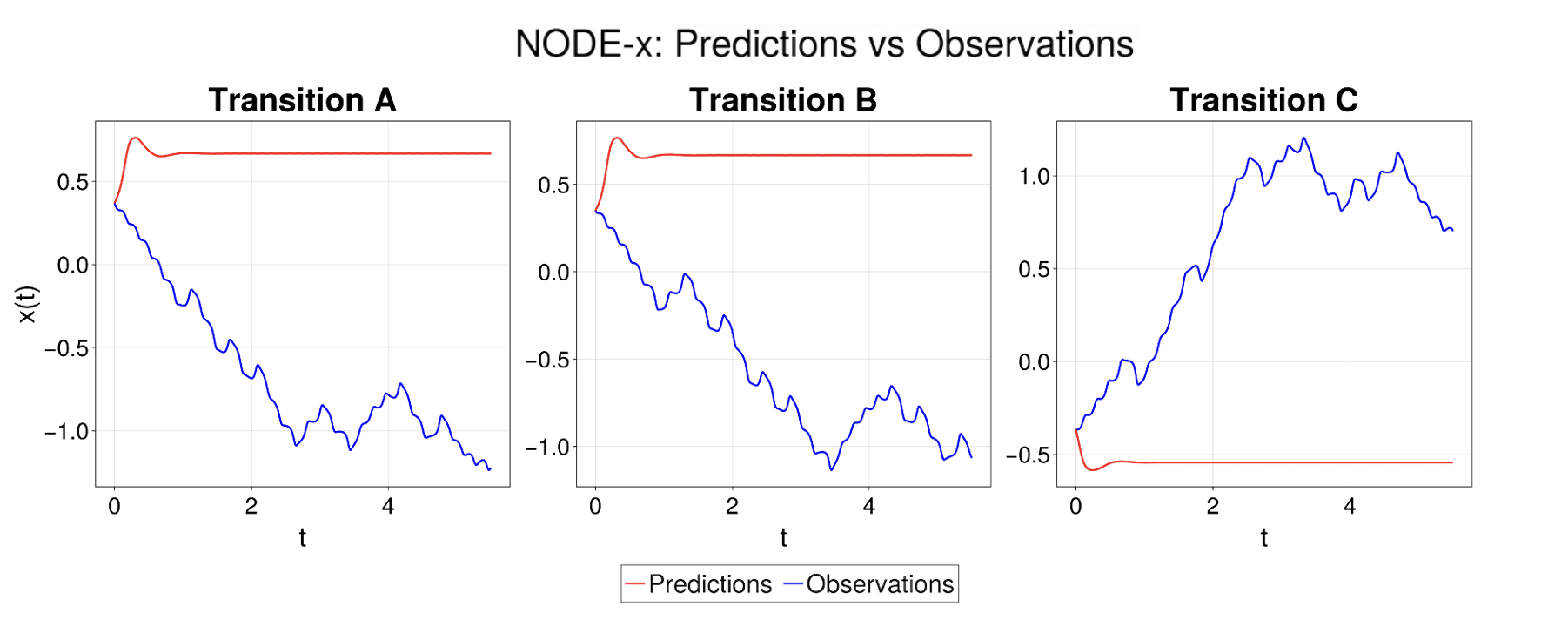}
    \caption{Forecasts for the same three transitions reported in Figure \ref{fig:full_forecast_overview} obtained using a NODE trained directly on the slow variable. The initial condition is chosen at lag $= 0.5$ before the transition.}
    \label{fig:NODEx}
\end{figure}

To benchmark our scale-separation approach, we trained a NODE directly on delay embeddings of $x(t)$ without decomposing slow and fast dynamics. Figure~\ref{fig:NODEx} shows that this direct approach fails to capture critical transitions, underscoring the necessity of explicitly modeling the fast chaotic forcing when slow and fast timescales become comparable near metastable transitions.

\newpage
\subsection{Bistable System with Multiplicative Chaotic Forcing}
\label{subsec:double_well_multiplicative}

The second system retains the same double-well potential but introduces state-dependent amplification of the chaotic forcing, a feature that significantly complicates both the statistical closure and the fast-dynamics surrogate. The slow-variable dynamics are governed by
\begin{equation} \label{eq:lorenz63_multiplicative}
\dot{x}(t) = x(t)\bigl[1 - x^2(t)\bigr] + \bigl[\sigma_1 + \sigma_2 x(t)\bigr]y_2(t),
\end{equation}
where $f(x) = x(1 - x^2) = -U'(x)$ remains unchanged from the first example, but the coupling function $g(x) = \sigma_1 + \sigma_2 x$ is now linear in $x$, yielding multiplicative noise. The fast subsystem follows Equation~\eqref{eq:general_coupled_system} with $\varepsilon = 0.5$, $\sigma_1 = 0.16$, and $\sigma_2 = 0.015$. Under the white-noise approximation, the effective Langevin equation takes the form $\dot{x}(t) = -U'(x) + \sqrt{2D_{\text{eff}}}[\sigma_1 + \sigma_2 x(t)]\,\xi(t)$, where $D_{\text{eff}} = \text{Var}[y_2] \cdot \tau_{\text{ACF}}$ is the effective diffusion coefficient (with $\tau_{\text{ACF}}$ the integral of the normalized ACF of $y_2(t)$). The presence of multiplicative noise fundamentally alters the relationship between the score function and the deterministic drift: the score no longer coincides with $-U'(x)/D_{\text{eff}}$ but instead reflects an effective potential that incorporates noise-induced corrections. For one-dimensional systems, the effective score can be derived analytically from the stationary Fokker--Planck equation under the Itô convention ($\sqrt{2}$ noise amplitude), yielding
\begin{equation}\label{eq:effective_score}
s_{\text{eff}}(x) = -\frac{U'(x)}{D_{\text{eff}}(\sigma_1 + \sigma_2 x)^2} - \frac{2\sigma_2}{\sigma_1 + \sigma_2 x}.
\end{equation}
The second term represents the noise-induced drift arising from the spatial variation of the diffusion coefficient; this contribution is absent in the additive case and must be captured by the data-driven score estimator to correctly reproduce the invariant measure. The stationary density in one dimension is $p(x) \propto [\Sigma^2(x)]^{-1} \exp\big(\int F(x)/\Sigma^2(x)\,dx\big)$ with $F(x) = -U'(x)$ the deterministic drift and $\Sigma^2(x) = D_{\text{eff}}(\sigma_1 + \sigma_2 x)^2$ the state-dependent diffusion coefficient; here we use the analytic score above as the benchmark.

\begin{figure}[H]
\centering
  \includegraphics[width=\linewidth]{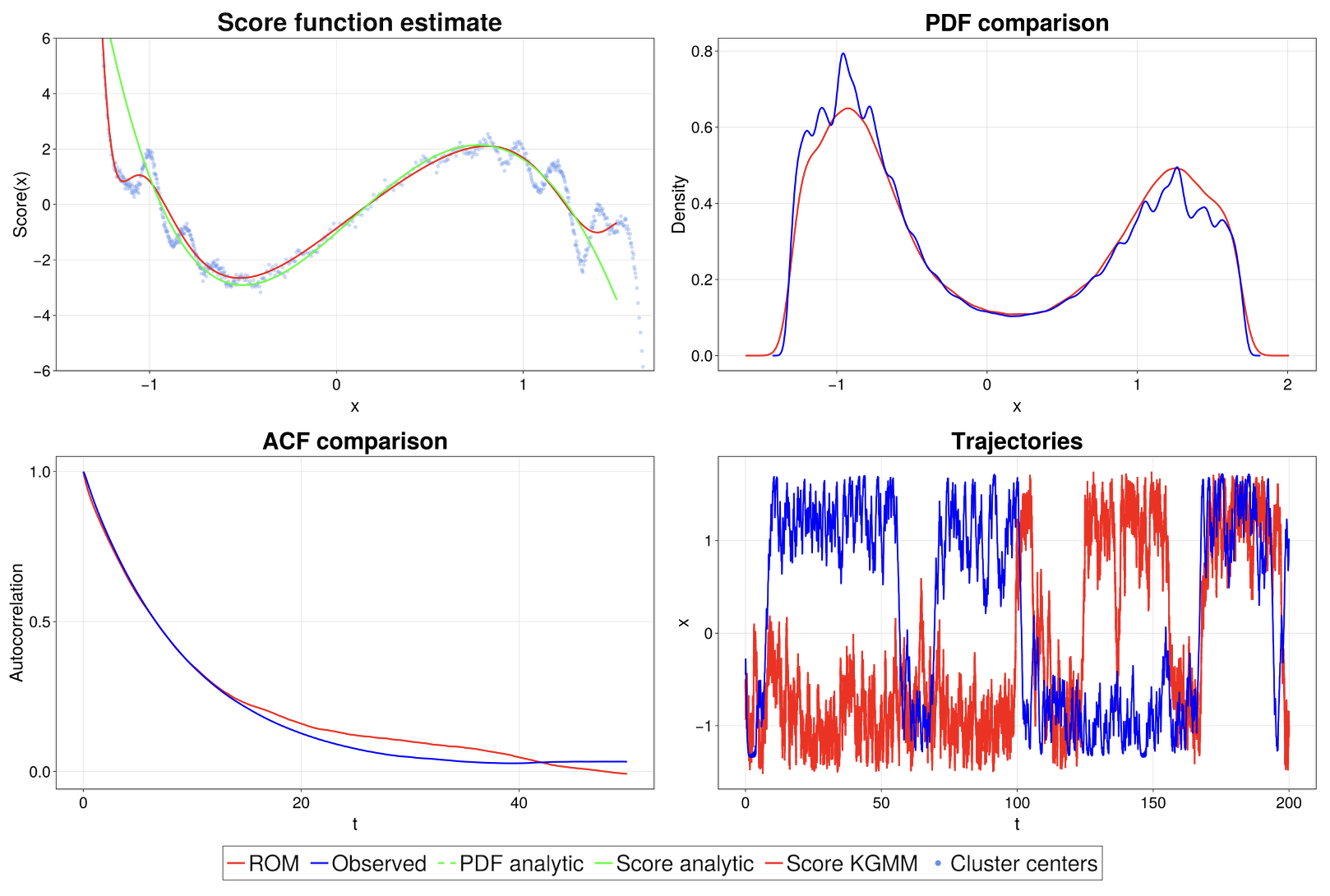}
\caption{Reduced-order model with Gaussian noise for system \ref{eq:lorenz63_multiplicative}. \textbf{Top row:} comparison between the stationary PDF reconstructed from the reduced-order model (red) and the empirical distribution of the original dataset (blue), alongside the KGMM-estimated score function and the analytic effective score for system \ref{eq:lorenz63_multiplicative}. \textbf{Bottom row:} comparison between autocorrelation functions of $x(t)$ obtained from the reduced-order model (red) and the original dataset (blue), together with trajectories from the stochastic reduced-order model (red) and the original system (blue) initialized identically.}
\label{fig:PDFACF_multiplicative}
\end{figure}

We apply KGMM score estimation and construct $\bm{\Phi}$ via moment matching as before. Figure~\ref{fig:PDFACF_multiplicative} demonstrates that the KGMM-estimated score (red) closely matches the analytical effective score (green). The top-left panel compares the steady-state PDF reconstructed from the ROM (red) with the empirical distribution from the full system (blue), showing excellent agreement. The ROM accurately reproduces both the PDF and the autocorrelation function (top-right and bottom panels) of the full system with high accuracy, confirming that the statistical closure correctly accounts for the noise-induced drift.

For short-term forecasting, the extracted fast signal $\widetilde{\vec{\xi}}(t)$ now exhibits pronounced heteroskedasticity due to the state-dependent coupling $g(x)$. Training a NODE directly on this heteroscedastic signal can destabilize gradient-based optimization, as minibatches sampled from different regions of state space exhibit vastly different variances. To address this, we apply the variance-normalization procedure detailed in Section~\ref{subsec:fast_variable_dynamics}: we train an auxiliary neural network with softplus output activation to learn the conditional variance $v_\theta(x) \approx \mathbb{E}[\widetilde{\vec{\xi}}^2 \mid x]$, minimizing the loss $\mathcal{L}(\theta) = (1/N)\sum_{i=1}^N (v_\theta(x_i) - r_i^2)^2$ over the training dataset, and then construct the gain function $\varphi(x) = \sqrt{v_\theta(x)}$ and the variance-normalized signal $\widetilde{\vec{\xi}}'(t) = \widetilde{\vec{\xi}}(t)/\varphi(x_t)$, which exhibits unit variance across the state space. The NODE is trained on delay-embedded vectors of $\widetilde{\vec{\xi}}'(t)$ with the same hyperparameters as in the additive case. During inference, the original heteroskedastic signal is reconstructed via $\widehat{\vec{\xi}}(t) = \varphi(x_t)\widehat{\vec{\xi}}'(t)$. The NODE's statistical properties and short-term trajectory accuracy are comparable to those achieved in the additive case (detailed in Appendix~\ref{appendix:NODE_performances}, Figure~\ref{fig:RMSE_node_short_multiplicative}).

\begin{figure}[H]
\centering
  \includegraphics[width=0.7\linewidth]{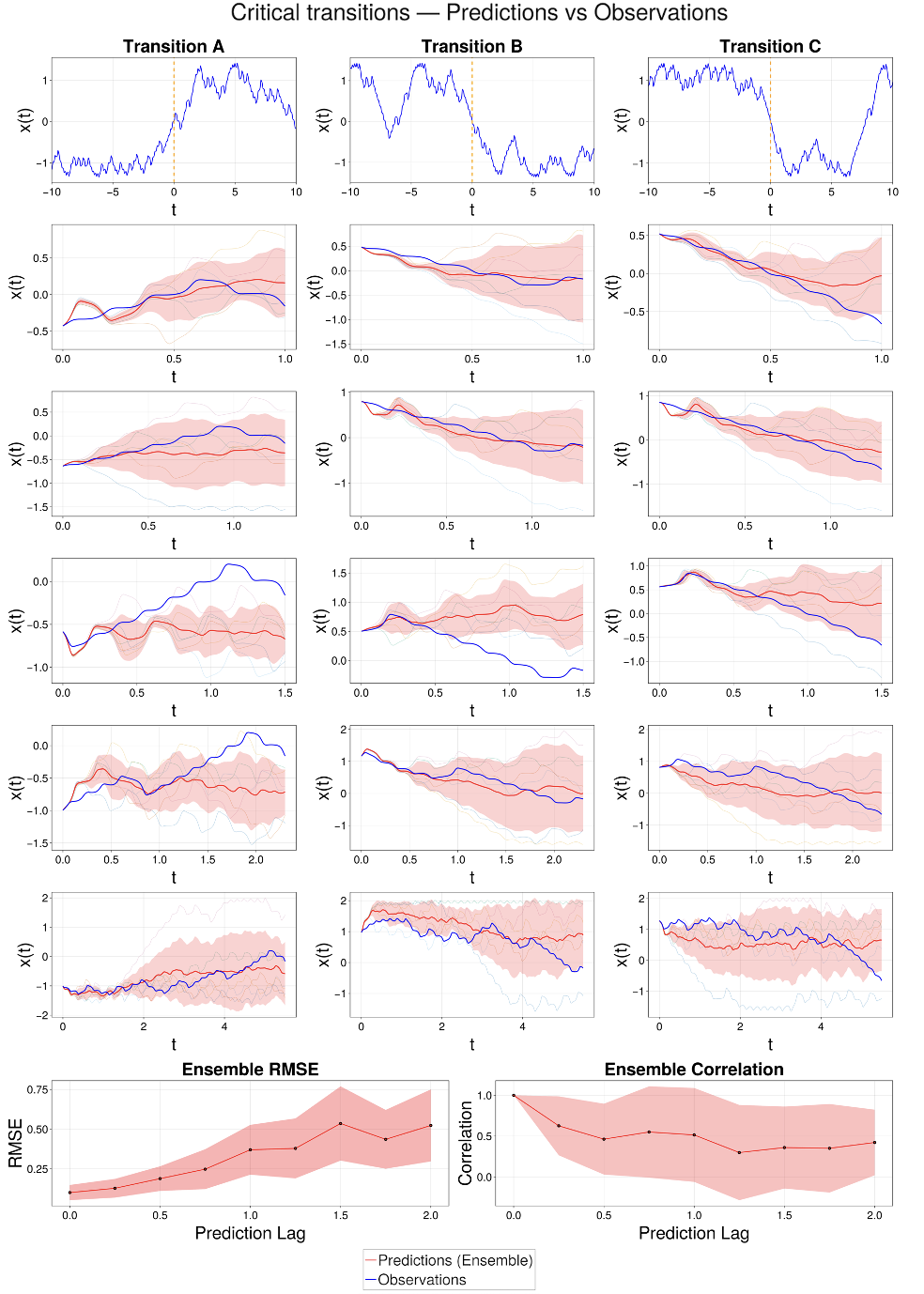}
\caption{\textbf{Reduced-order forecasts with Neural ODEs and error growth.} \textbf{Top:} ground truth (blue) vs.\ ensemble mean of five Neural ODEs (red) with $\pm 1\sigma$ spread (shaded). \textbf{Bottom:} ensemble-mean RMSE vs.\ lead time with variability band; beyond a lead time of $\sim 1$ the RMSE exceeds $0.5$ and the correlation drops below $0.5$, indicating loss of predictive power.}
\label{fig:CriticalTransMult}
\end{figure}

We evaluate the hybrid model's transition-forecasting capability using an ensemble of five independently trained NODE instances. Figure~\ref{fig:CriticalTransMult} presents the ensemble-mean forecasts (red, with $\pm 1\sigma$ spread) versus ground truth (blue) for three representative transitions, along with RMSE as a function of lead time. The results are consistent with the additive case: the model retains non-trivial predictive skill (RMSE $< 0.5$, correlation $> 0.5$) up to approximately $t = 1$ lead time, successfully anticipating critical transitions within this horizon. The variance-normalization strategy proves essential for achieving stable NODE training and accurate short-horizon forecasts in the presence of state-dependent forcing.

\newpage

\subsection{Tristable System with Slow Cycloperiodic Forcing}
\label{subsec:tristable_slow_periodic}

The third system introduces non-autonomous dynamics through explicit time-dependence in the deterministic drift, combined with a tristable potential landscape. The slow-variable dynamics are governed by
\begin{equation}
\label{eq:periodic_syst}
\dot{x}(t) = -U'\!\big(x(t)\big) + A\cos(\omega t) + \sigma_0\,y_2(t),
\end{equation}
where the deterministic component $f(x,t) = -U'(x) + A\cos(\omega t)$ combines a tristable potential
\begin{equation}
\label{eq:tristable_potential}
U(x) = \frac{50}{131}\!\left(\frac{x^{6}}{3}-\frac{122}{64}\,x^{4}+\frac{79}{29}\,x^{2}\right)+\frac{1}{50}\,x^{2},
\quad
U'(x) = \frac{50}{131}\!\left(2x^{5}-\frac{61}{8}\,x^{3}+\frac{158}{29}\,x\right)+\frac{1}{25}\,x
\end{equation}
with local minima at $x = \pm 1.25$ and $x = 0$, and a sinusoidal forcing term with amplitude $A = 0.5$ and angular frequency $\omega = 2\pi \cdot 0.002$, corresponding to period $T = 500$. The coupling function $g(x) = \sigma_0 = 0.08$ is constant (additive forcing), and the fast subsystem follows Equation~\eqref{eq:general_coupled_system} with $\varepsilon = 0.5$. Since the period $T \gg t_{\text{fast}}$ (where $t_{\text{fast}}$ is the decorrelation time of the Lorenz 63 attractor), the cycloperiodic component evolves on the slow timescale and must be incorporated into the statistical closure rather than treated as part of the fast forcing.

To handle the non-autonomous dynamics, we employ the state-space augmentation strategy detailed in Section~\ref{subsubsec:cyclostationary_extension}: we augment the physical state $x(t)$ with harmonic clock variables $(\sin(\omega t), \cos(\omega t))$ to form the three-dimensional augmented state $\vec{x}_{\text{aug}}(t) = (\sin(\omega t), \cos(\omega t), x(t)) \in \mathbb{R}^3$, thereby recasting the non-autonomous problem as an autonomous one in the augmented space. Applying KGMM to the augmented time series yields a score function $\vec{s}_{\text{aug}}(\vec{x}_{\text{aug}})$ in $\mathbb{R}^3$, from which the time-dependent physical score is recovered by evaluating $\vec{s}_{\text{aug}}$ at the current clock phase: $\vec{s}(x,t) = \vec{s}_{\text{aug}}(\sin(\omega t), \cos(\omega t), x)$. This time-dependent score can be compared with the analytical expression derived from the deterministic drift: $\vec{s}_{\text{an}}(x,t) = \bigl[-U'(x) + A\cos(\omega t)\bigr]/D_{\text{eff}}$, where $D_{\text{eff}}$ is the effective diffusion coefficient. The drift tensor $\bm{\Phi}_{\text{aug}}$ is constructed via moment matching in the augmented space, and the physical-coordinate block is extracted to define the time-dependent ROM.

\begin{figure}[H]
\centering
\includegraphics[width=\linewidth]{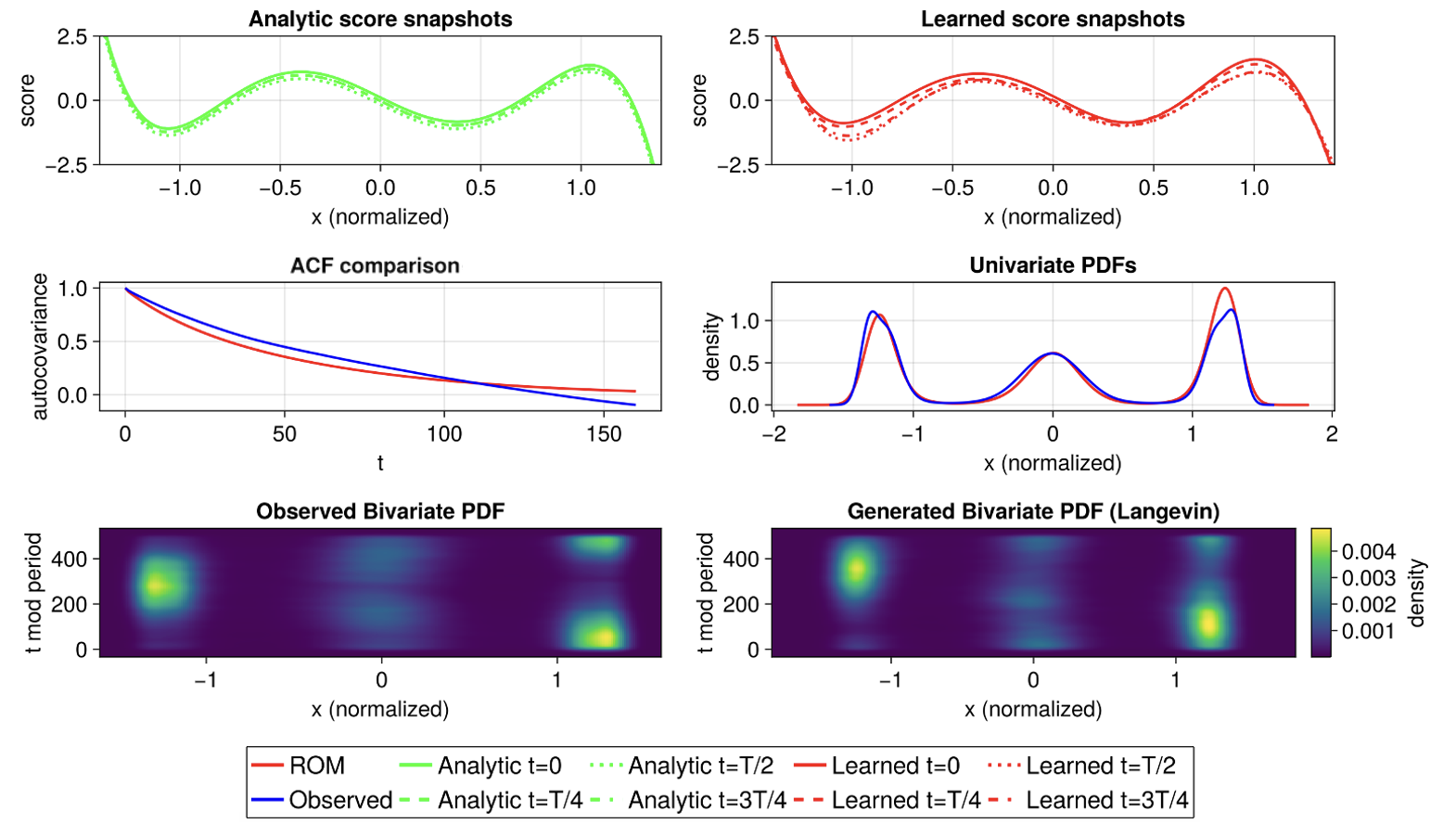}
\caption{\textbf{Top panel:} time-dependent score function evaluated at four phases \(t\in\{0,\,T/4,\,T/2,\,3T/4\}\): KGMM estimate (right, red) vs.\ analytic expression \(-U'(x)+A\cos(\omega t)\) (left, green). \textbf{Central panel:} cycle-averaged marginal PDF (left) and autocorrelation function (right) of \(x(t)\): reduced-order model (red) vs.\ empirical estimates from the observed series (blue) for the periodically forced system \eqref{eq:periodic_syst}. \textbf{Bottom panel:} phase-resolved joint PDF \(\rho(x,\varphi)\), with \(\varphi=\omega t \bmod 2\pi\): empirical estimate (left) and reconstruction from the reduced-order model \eqref{eq:effective_dynamics_short} (right).}
\label{fig:UPDFeACFperiodic}
\end{figure}

Figure~\ref{fig:UPDFeACFperiodic} demonstrates the accuracy of the augmented KGMM approach: the top panel shows that the estimated time-dependent score (red) closely matches the analytical expression (green) at four representative phases $t \in \{0, T/4, T/2, 3T/4\}$, capturing both the potential gradient and the sinusoidal forcing contribution. The central panels confirm statistical consistency by comparing the cycle-averaged marginal PDF (left) and autocorrelation function (right) of the ROM (red) against empirical estimates from the full system (blue), showing excellent agreement between the two. Because the system is non-autonomous, we also assess the phase-resolved joint PDF $\rho(x, \varphi)$ with $\varphi = \omega t \bmod 2\pi$ (bottom panel): the ROM accurately reconstructs the empirical phase-space density, including the time-dependent shifts in the relative populations of the three wells induced by the periodic forcing.

For short-term forecasting, the fast subsystem remains the Lorenz 63 attractor without explicit time dependence, so the NODE is trained with the same architecture and hyperparameters as in the previous examples. The NODE surrogate reproduces the statistical properties and short-term trajectories of $y_2(t)$ with comparable accuracy (details in Appendix~\ref{appendix:NODE_performances}, Figure~\ref{fig:NODEperformance_periodic}).

\begin{figure}[H]
\centering
  \includegraphics[width=0.7\linewidth]{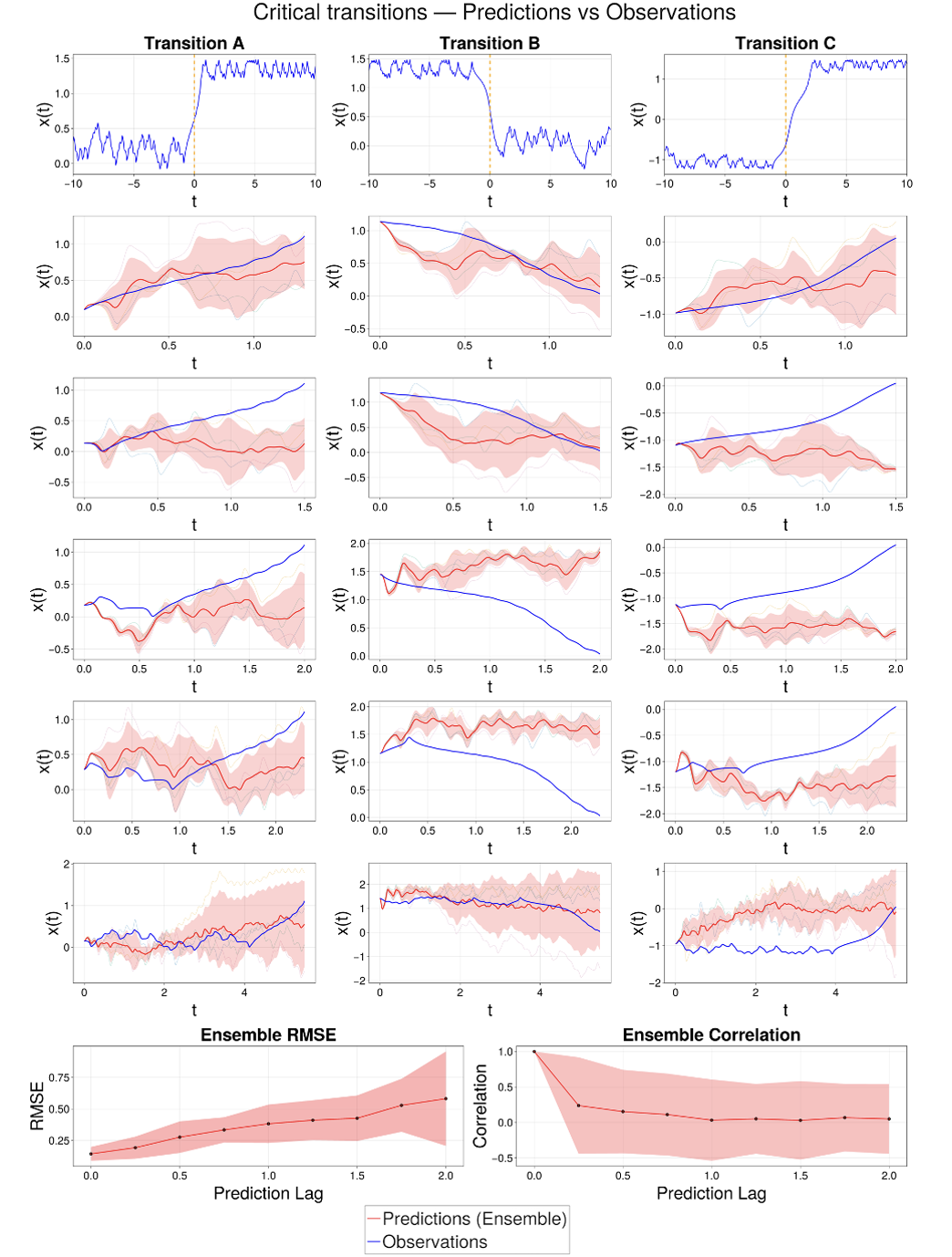}
\caption{\textbf{Top:} ensemble forecasts (red, mean \(\pm 1\sigma\)) of the reduced-order model with NODE fast forcing compared with the observed trajectory (blue) for multiple lead times before a critical transition; \textbf{Bottom:} ensemble-mean RMSE as a function of prediction lag, with the shaded band indicating inter-model variability.}
\label{fig:ensemble_periodic_forecast}
\end{figure}

We evaluate the hybrid model's ability to forecast critical transitions among the three metastable wells by initializing ensemble forecasts (averaged over five NODE instances) at multiple lead times before observed transitions. Figure~\ref{fig:ensemble_periodic_forecast} presents representative results: the ensemble-mean forecast (red, with $\pm 1\sigma$ spread) versus ground truth (blue) for varying lead times, along with RMSE as a function of forecast horizon. Despite the additional complexity introduced by non-autonomous dynamics and the tristable landscape, the results remain consistent with the bistable cases: the ROM retains non-trivial predictive power up to lead times comparable to those achieved in the bistable systems, successfully anticipating transitions within this horizon. The state-space augmentation approach proves effective for handling cycloperiodic forcing when the period is long relative to the fast-process decorrelation time.

\newpage
\subsection{Tristable System with Fast Cycloperiodic Forcing}
\label{subsec:tristable_fast_periodic}

The fourth and final system examines the complementary regime in which the cycloperiodic forcing operates on a timescale comparable to the fast-process decorrelation time. The slow-variable dynamics retain the same form as in the third example,
\begin{equation}
\label{eq:periodic_syst_fast}
\dot{x}(t) = -U'\!\big(x(t)\big) + A\cos(\omega t) + \sigma_0\,y_2(t),
\end{equation}
with the same tristable potential $U(x)$ and parameters $A = 0.5$, $\sigma_0 = 0.08$, and $\varepsilon = 0.5$, but now the angular frequency is increased to $\omega = 2\pi$, corresponding to period $T = 1$. Since $T \sim t_{\text{fast}}$, the cycloperiodic term evolves on the same timescale as the chaotic Lorenz 63 attractor and must therefore be incorporated into the fast forcing rather than the slow statistical closure. This regime poses a distinct modeling challenge: the long-term statistical properties are governed solely by the time-averaged dynamics $\dot{x} = -U'(x) + \sigma_0 y_2(t)$ (as the sinusoidal term has zero mean over full cycles), while short-term forecasts must account for the explicit time-dependence of the combined fast forcing.

\begin{figure}[H]
\centering
  \includegraphics[width=\linewidth]{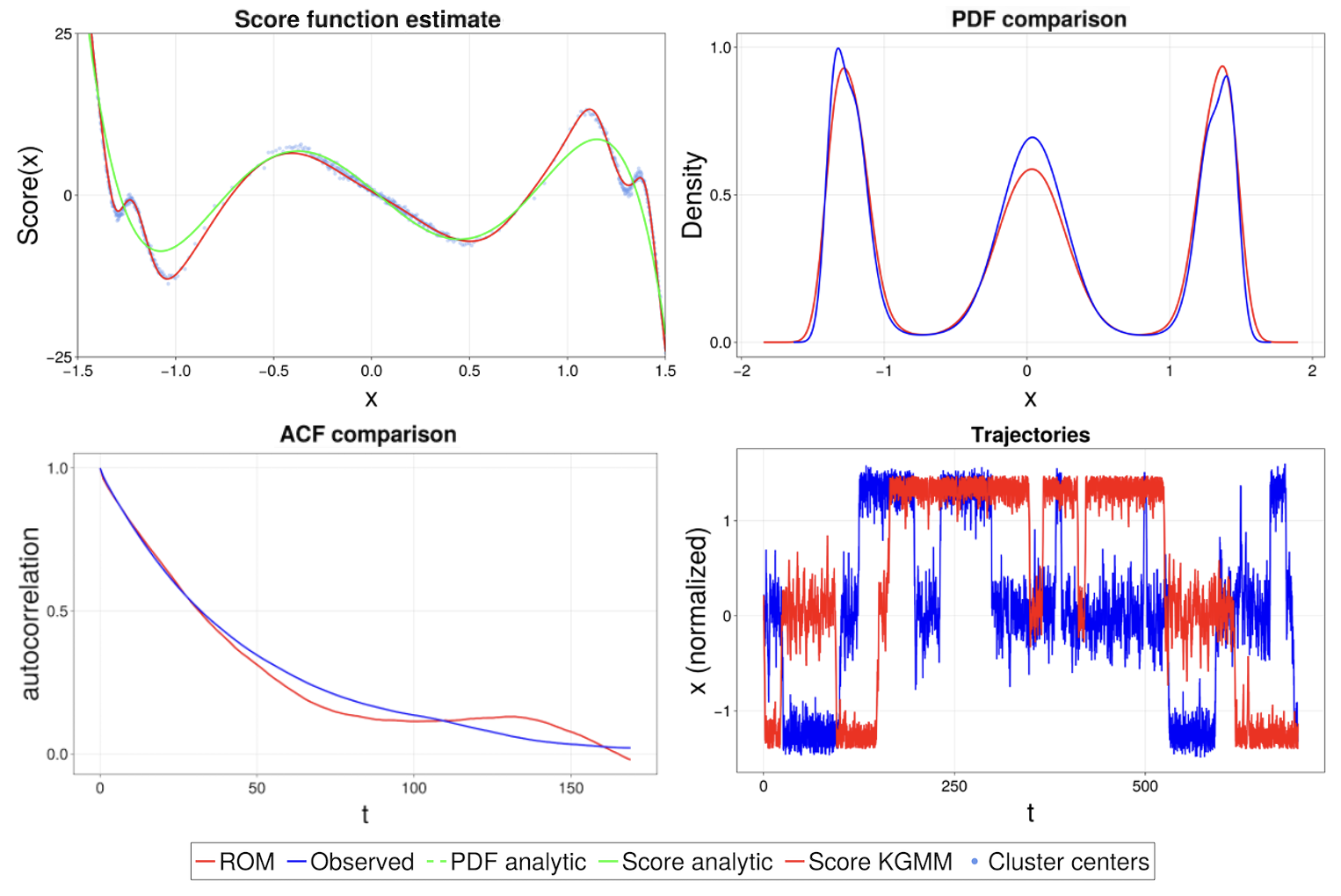}
\caption{\textbf{Top row:} score function estimated via KGMM (solid) compared with the analytic derivative \(U'(x)\) (left) and stationary PDF comparison (right): reduced-order model (red) vs.\ empirical distribution from the observed series (blue) for the system \ref{eq:periodic_syst} with \(T=1\). \textbf{Bottom row:} autocorrelation functions (right) of \(x(t)\) and trajectories for the periodically forced system \ref{eq:periodic_syst} with fast periodic term, contrasting the stochastic reduced-order model (red) with the observed series (blue).}
\label{fig:PDFeACFperiodicfast}
\end{figure}

For the statistical closure, we apply KGMM to the physical state $x(t)$ without augmentation, treating the system as effectively autonomous for long-term statistics. The resulting score function $\vec{s}(x)$ corresponds to the time-independent component $-U'(x)/D_{\text{eff}}$, where $D_{\text{eff}}$ is the effective diffusion coefficient determined by the fast-process decorrelation time. The drift tensor $\bm{\Phi}$ is constructed via moment matching, and the ROM $\dot{x} = \bm{\Phi}\,\vec{s}(x) + \sqrt{2}\,\text{chol}(\bm{\Phi}_S)\,\xi(t)$ is validated by comparing the cycle-averaged (time-invariant) steady-state PDF and autocorrelation function against empirical statistics from the full system. Figure~\ref{fig:PDFeACFperiodicfast} confirms that the ROM accurately reproduces the long-term statistical properties despite neglecting the fast periodic component in the closure.

For short-term forecasting, the extracted fast signal $\widetilde{\vec{\xi}}(t)$ now contains contributions from both the chaotic Lorenz 63 process and the sinusoidal forcing term. To enable the NODE to learn this combined forcing, we augment each delay-embedded vector $\mathbf{z}'(t) = [\widetilde{\vec{\xi}}'(t), \widetilde{\vec{\xi}}'(t - \Delta\tau), \ldots, \widetilde{\vec{\xi}}'(t - (m-1)\Delta\tau)]$ with harmonic clock variables $(\sin(\omega t), \cos(\omega t))$ to form $\mathbf{z}'_{\text{aug}}(t) = [(\sin(\omega t), \cos(\omega t)), \mathbf{z}'(t)] \in \mathbb{R}^{2 + m}$, as described in Section~\ref{subsec:fast_variable_dynamics}. The NODE is trained to predict $\frac{d}{dt}{\widetilde{\vec{\xi}}}'(t) = \vec{G}_\theta(\mathbf{z}'_{\text{aug}}(t))$ with the augmented input, enabling it to recognize and reproduce the periodic modulation of the fast signal. During inference, the clock variables are evaluated at the current simulation time, providing the NODE with the instantaneous phase of the sinusoidal forcing. Figure~\ref{fig:NODEperformancePeriodicFast} in Appendix~\ref{appendix:NODE_performances} demonstrates that the augmented NODE accurately reproduces both the statistical properties (PDF and ACF) and short-term trajectories of the observed fast signal, including the periodic modulation.

\begin{figure}[H]
\centering
  \includegraphics[width=0.7\linewidth]{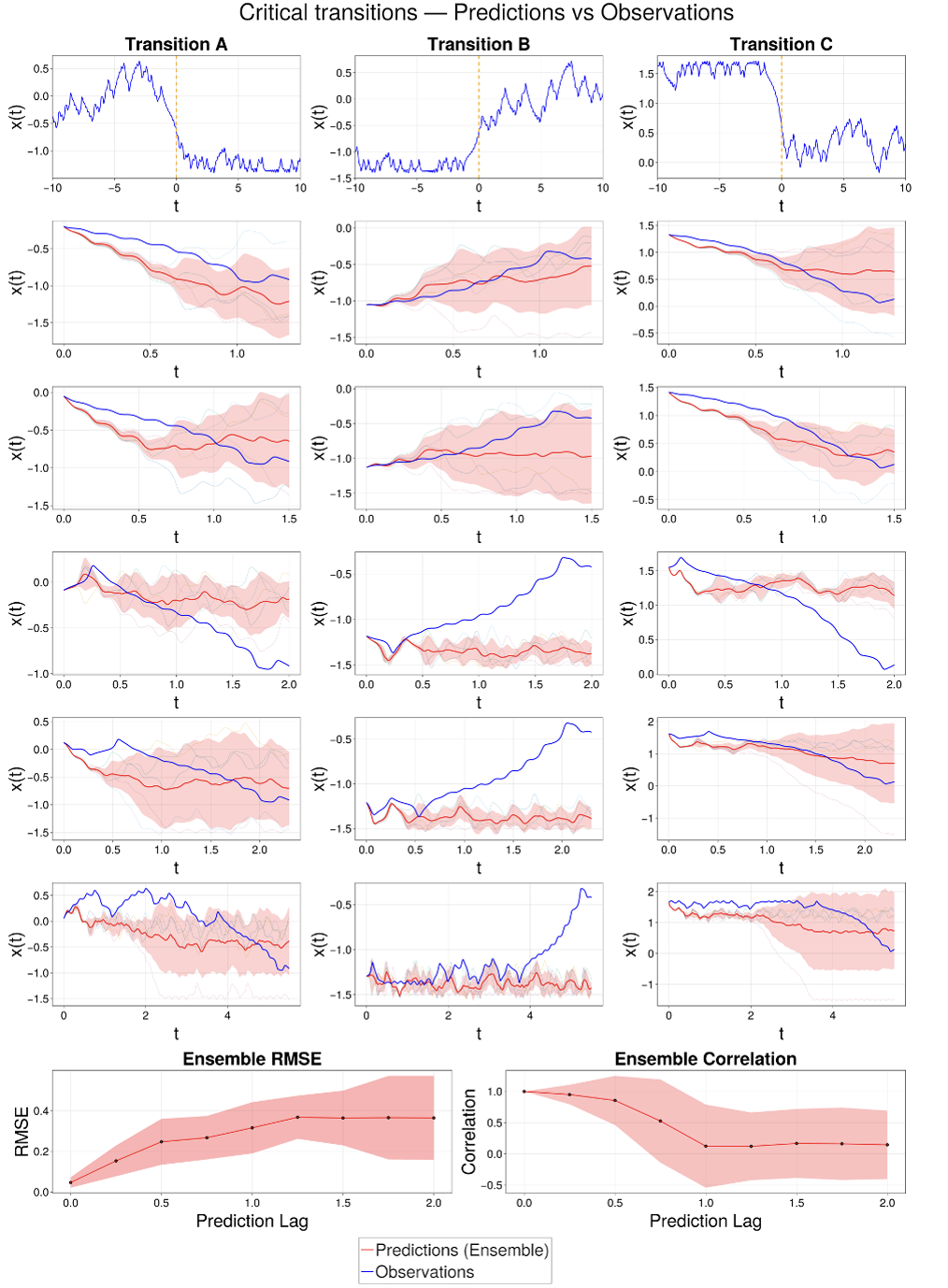}
\caption{\textbf{Top:} ensemble forecasts (red, mean \(\pm 1\sigma\)) of the ROM with NODE fast forcing vs.\ the observed trajectory (blue) for multiple lead times prior to a critical transition. \textbf{Bottom:} ensemble-mean RMSE vs.\ prediction lag. The shaded band denotes inter-model variability.}
\label{fig:ensemble_periodic_forecast_fast}
\end{figure}

We evaluate the hybrid model's transition-forecasting capability using an ensemble of five independently trained NODE instances, initializing forecasts at multiple lead times before observed critical transitions. Figure~\ref{fig:ensemble_periodic_forecast_fast} presents the ensemble-mean forecast (red, with $\pm 1\sigma$ spread) versus ground truth (blue), along with RMSE as a function of forecast horizon. The results are consistent with the previous systems, though the predictive horizon is slightly extended: the ROM retains non-trivial predictive skill (RMSE $< 0.5$, correlation $> 0.5$) up to lead times slightly longer than those achieved in the other examples, successfully anticipating critical transitions within this window. The marginally improved performance may reflect the additional information provided by the deterministic sinusoidal component, which partially constrains the fast forcing and reduces effective stochasticity. This final example demonstrates that the hybrid framework can accommodate cycloperiodic forcing across both slow and fast timescale regimes through appropriate assignment of the periodic component to either the statistical closure (via state-space augmentation) or the fast-dynamics surrogate (via delay-vector augmentation), depending on the relationship between the forcing period and the fast-process decorrelation time.

\newpage 
\section{Conclusions}
\label{sec:conclusions}

We have presented a hybrid data-driven framework for reduced-order modeling of multiscale dynamical systems that integrates score-based statistical closures with Neural Ordinary Differential Equation surrogates for fast chaotic drivers. The approach addresses a fundamental challenge in the modeling of complex systems exhibiting timescale separation: preserving both the long-term statistical fidelity necessary for capturing invariant measures and the short-term dynamical accuracy required for anticipating rare critical transitions between metastable states. By combining these two complementary objectives within a unified architecture, our methodology extends beyond conventional Gaussian noise approximations and enables quantitatively accurate predictions even when slow and fast timescales become comparable near transition events.

The framework comprises two principal components, each designed to capture distinct aspects of the multiscale dynamics. For the slow-variable dynamics, we employ the K-means Gaussian Mixture Model (KGMM) method to estimate the score function—the gradient of the logarithm of the steady-state probability density—directly from time-series observations, without requiring knowledge of the governing equations or the normalizing constant. This learned score function is embedded into the drift term of an effective Langevin equation via a data-driven construction of the drift and diffusion tensors through moment matching, ensuring that the reduced-order model (ROM) reproduces the observed invariant measure by construction. For systems with explicit time dependence, we introduced a state-space augmentation strategy that incorporates harmonic clock variables to handle cycloperiodic forcing, enabling score-based closures for non-autonomous dynamics. For the fast-variable dynamics, we move beyond the standard Gaussian colored-noise approximation by training a Neural ODE on delay-embedded vectors of the extracted fast signal. This continuous-time surrogate captures the chaotic, non-Gaussian features of the fast forcing with sufficient fidelity to enable accurate short-horizon forecasts. In systems exhibiting state-dependent amplification of the fast forcing (multiplicative noise), we developed a variance-normalization procedure that stabilizes NODE training by homogenizing the conditional variance of the fast signal across state space.

We validated this hybrid methodology on a hierarchy of four prototypical one-dimensional metastable systems driven by the Lorenz 63 chaotic attractor, systematically exploring scenarios of increasing complexity: bistable potentials with additive and multiplicative forcing, and tristable non-autonomous systems with cycloperiodic components operating on both slow and fast timescales relative to the decorrelation time of the chaotic driver. Across all four systems, the score-based ROM with Gaussian white-noise forcing accurately reproduced the empirical steady-state probability density functions and autocorrelation functions over extended time horizons, demonstrating statistical consistency. Where analytical benchmarks were available, the KGMM estimates exhibited excellent agreement, validating the data-driven closure procedure. For the bistable additive case, we compared against both the analytical score function and the analytical invariant measure derived under the Gaussian white-noise approximation; the slight deviations between the analytical and observed PDFs confirm the non-Gaussian nature of the chaotic forcing, while the data-driven ROM accurately captures the true empirical statistics. For the multiplicative case, we validated against the analytical effective score incorporating noise-induced drift. The NODE surrogates for the fast chaotic forcing reproduced both the statistical properties (PDF and ACF) and short-term trajectory predictions up to approximately the Lyapunov time of the Lorenz 63 subsystem, confirming the approach's ability to capture complex nonlinear dynamics without explicit knowledge of the governing equations.

Most significantly, the hybrid ROM combining score-based drift with NODE-generated fast forcing successfully anticipated critical transitions between metastable wells with lead times approaching the Lyapunov time of the chaotic driver. Ensemble forecasts (averaged over multiple independently trained NODE instances) exhibited non-trivial predictive skill, as quantified by root mean square error and temporal correlation, for forecast horizons extending beyond those achievable by either the purely statistical closure or direct NODE modeling of the slow variable alone. This capability to resolve rare transition events distinguishes the hybrid framework from conventional approaches: by explicitly modeling the intermittency, memory, and state-dependent amplification of the fast chaotic forcing rather than replacing it with Gaussian noise, the method retains sufficient dynamical information to capture the transient behaviors that trigger escape from metastable basins of attraction. Benchmark comparisons with a NODE trained directly on the slow-variable time series—without decomposing slow and fast dynamics—demonstrated that such direct approaches fail to anticipate critical transitions, underscoring the necessity of scale separation when timescales become comparable near metastable boundaries.

The entirely data-driven nature of our framework—learning both the statistical closure and the fast-dynamics surrogate from observations alone—represents a key advantage over previous methodologies \cite{lim2020predicting} that assume knowledge of the analytic form of the slow-variable dynamics. This feature positions the hybrid ROM as a practical tool for systems where governing equations are unknown, partially known, or computationally intractable to resolve directly. The flexibility of the approach extends to diverse scenarios: additive versus multiplicative coupling of fast and slow variables, autonomous versus non-autonomous dynamics with cycloperiodic forcing, and regimes where the forcing period is either much longer or comparable to the fast-process decorrelation time. The state-space and delay-vector augmentation strategies introduced for handling explicit time dependence provide a principled mechanism for restoring effective autonomy, enabling the application of autonomous learning algorithms (KGMM for score estimation, delay-embedding NODEs for fast-signal surrogate construction) to inherently non-stationary problems.

Several directions for future research emerge from this work. First, extension to higher-dimensional systems—both in the dimensionality of the slow variables and the complexity of the fast chaotic drivers—will require scalable score-estimation techniques and careful consideration of the curse of dimensionality in delay-embedding constructions. Recent advances in denoising score matching and continuous normalizing flows may offer computationally tractable alternatives to KGMM for high-dimensional score learning, while dimension-reduction techniques (e.g., autoencoders, Koopman mode decomposition) could be integrated into the delay-embedding pipeline to mitigate computational costs. Second, real-world applications often involve partial, noisy observations of the slow variables, necessitating data assimilation strategies to infer unobserved state components and uncertainty quantification frameworks to assess forecast reliability. Third, the assumption of stationary fast dynamics—inherent in the delay-embedding approach—may be violated in non-stationary systems exhibiting regime shifts, seasonal modulation, or gradual parameter drift; adaptive or time-varying NODE architectures could address such scenarios. Fourth, while our validation focused on systems with known ground truth to enable rigorous benchmarking, application to observational data from climate, turbulence, or neuroscience will require careful validation protocols and physical interpretability constraints to ensure that learned models respect conservation laws, symmetries, or other domain-specific priors.

In conclusion, the hybrid framework presented here establishes a principled methodology for combining statistical closure techniques with explicit surrogate models of fast chaotic dynamics, offering a pathway toward predictive modeling of complex multiscale phenomena where both long-term statistical properties and short-term transient behaviors are essential. By demonstrating that data-driven score-based ROMs augmented with NODE surrogates can anticipate rare critical transitions in metastable systems driven by chaotic forcing, this work opens new possibilities for early-warning systems in climate tipping points, phase transitions in materials, regime shifts in ecosystems, and other high-impact applications where timely prediction of abrupt changes remains a central challenge.

\appendix
\section*{Appendix}
\section{Effective White Noise Approximation}
\label{sec:appendix_white_noise}

Here we present a rigorous derivation showing how colored noise processes can be approximated by effective white noise when dynamics are observed on time scales exceeding the noise correlation time. This mathematical framework provides the foundation for the approximations used in the main text.

Consider a multivariate stochastic process governed by an overdamped Langevin equation:
\begin{equation}
\dot{\vec{x}}(t) = \vec{F}(\vec{x}(t)) + \bm{\Sigma}_0(\vec{x}(t))\vec{\eta}(t),
\end{equation}
where $\vec{\eta}(t)$ is a colored noise process with zero mean and component-wise exponentially decaying correlations:
\begin{equation}
\langle \eta_i(t) \eta_j(t') \rangle = \delta_{ij} e^{-|t-t'|/\tau_i},
\end{equation}
where $\tau_i$ is the characteristic correlation time for the $i$-th noise component.

Our objective is to find an effective white noise representation that preserves the statistical properties of the system when observed at time scales much larger than any $\tau_i$. To accomplish this, we analyze the displacement over a time interval $[0,t]$:
\begin{equation}
\vec{x}(t) - \vec{x}(0) = \int_0^t \vec{F}(\vec{x}(s))\,ds + \int_0^t \bm{\Sigma}_0(\vec{x}(s))\vec{\eta}(s)\,ds.
\end{equation}

For sufficiently small time intervals where $\bm{\Sigma}_0(\vec{x})$ can be treated as approximately constant, we focus on the noise contribution to the mean-square displacement for each component:
\begin{equation}
\left\langle \left( \int_0^t [\bm{\Sigma}_0\vec{\eta}(s)]_i\,ds \right)^2 \right\rangle = \sum_{j,k} \int_0^t \int_0^t \Sigma_{0,ij} \Sigma_{0,ik} \langle \eta_j(s)\eta_k(s') \rangle \,ds\,ds'.
\end{equation}

Using the correlation structure of $\vec{\eta}$:
\begin{equation}
\left\langle \left( \int_0^t [\bm{\Sigma}_0\vec{\eta}(s)]_i\,ds \right)^2 \right\rangle = \sum_j \int_0^t \int_0^t \Sigma_{0,ij}^2 e^{-|s-s'|/\tau_j}\,ds\,ds'.
\end{equation}

For time scales $t \gg \max(\tau_j)$, each double integral asymptotically approaches:
\begin{equation}
\int_0^t \int_0^t e^{-|s-s'|/\tau_j}\,ds\,ds' \approx 2\tau_j t,
\end{equation}
leading to:
\begin{equation}
\left\langle \left( \int_0^t [\bm{\Sigma}_0\vec{\eta}(s)]_i\,ds \right)^2 \right\rangle \approx 2t \sum_j \Sigma_{0,ij}^2 \tau_j.
\end{equation}

For white noise $\vec{\xi}(t)$ with $\langle \xi_i(t)\xi_j(t') \rangle = \delta_{ij}\delta(t-t')$, the corresponding expression would be:
\begin{equation}
\left\langle \left( \int_0^t [\bm{A}\vec{\xi}(s)]_i\,ds \right)^2 \right\rangle = t \sum_j A_{ij}^2.
\end{equation}

To match these statistical properties, we define a new noise-coupling matrix:
\begin{equation}
\Sigma_{ij} = \Sigma_{0,ij}\sqrt{2\tau_j},
\end{equation}
which can be written in matrix form as:
\begin{equation}
\bm{\Sigma} = \bm{\Sigma}_0 \cdot \text{diag}(\sqrt{2\bm{\tau}}).
\end{equation}


This approximation becomes increasingly accurate as the ratio of observational time scale to the largest correlation time grows, allowing us to replace the original dynamics with colored noise by an effective dynamics with appropriately scaled white noise. The componentwise scaling by $\sqrt{2\tau_j}$ emerges naturally as the necessary adjustment to preserve the statistical properties of the process when transitioning from colored to white noise.\newpage

\section{NODE Performances}
\label{appendix:NODE_performances}
\subsection{NODE performances for Example II}
\begin{figure}[H]
    \centering
        \includegraphics[width=1\linewidth]{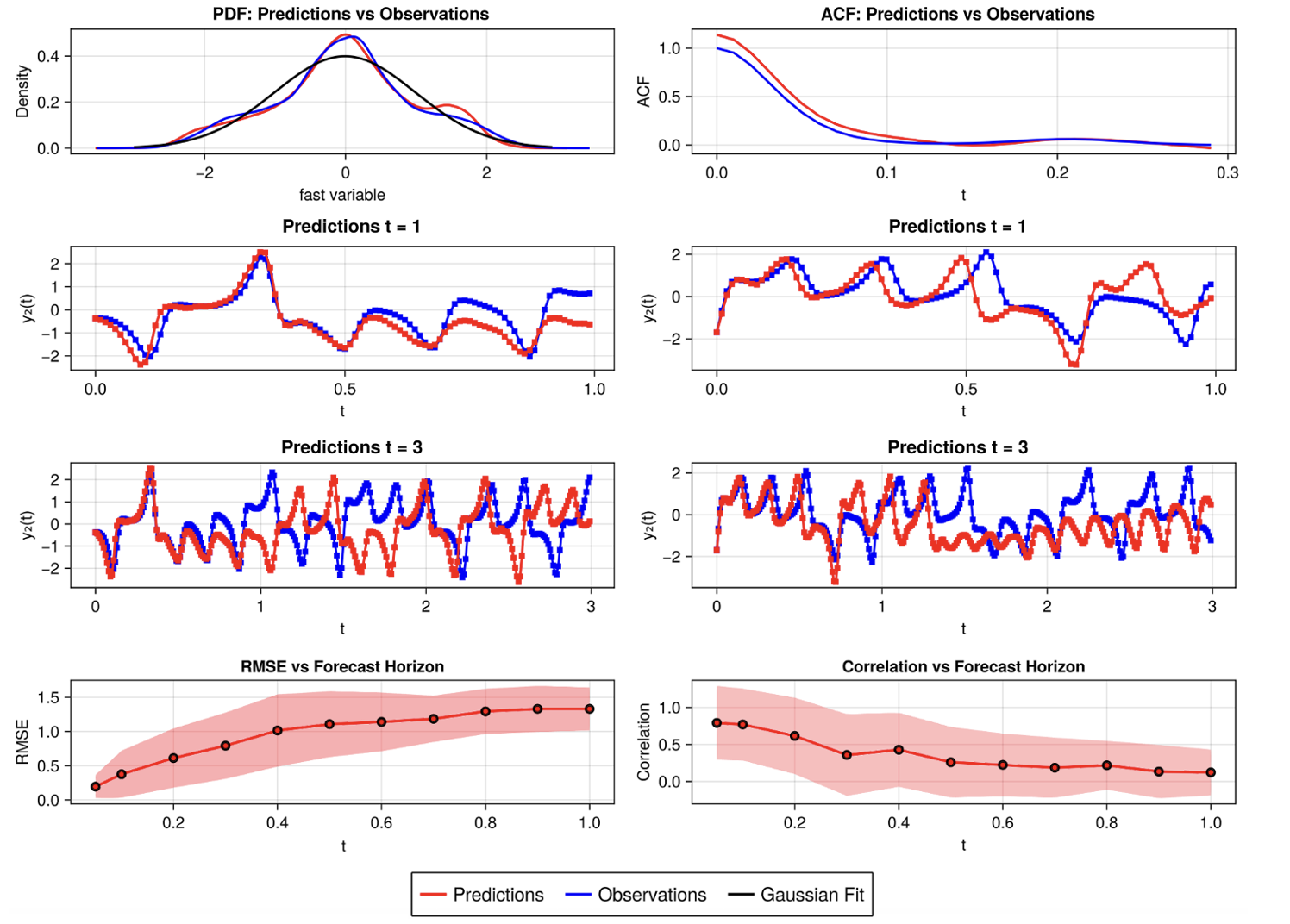}
    \caption{Example II, Performances of the NODE for system \ref{eq:lorenz63_multiplicative}}
    \label{fig:RMSE_node_short_multiplicative}
\end{figure}

\subsection{NODE performances for Example IIIa}
\begin{figure}[H]
    \centering
    \includegraphics[width=1\linewidth]{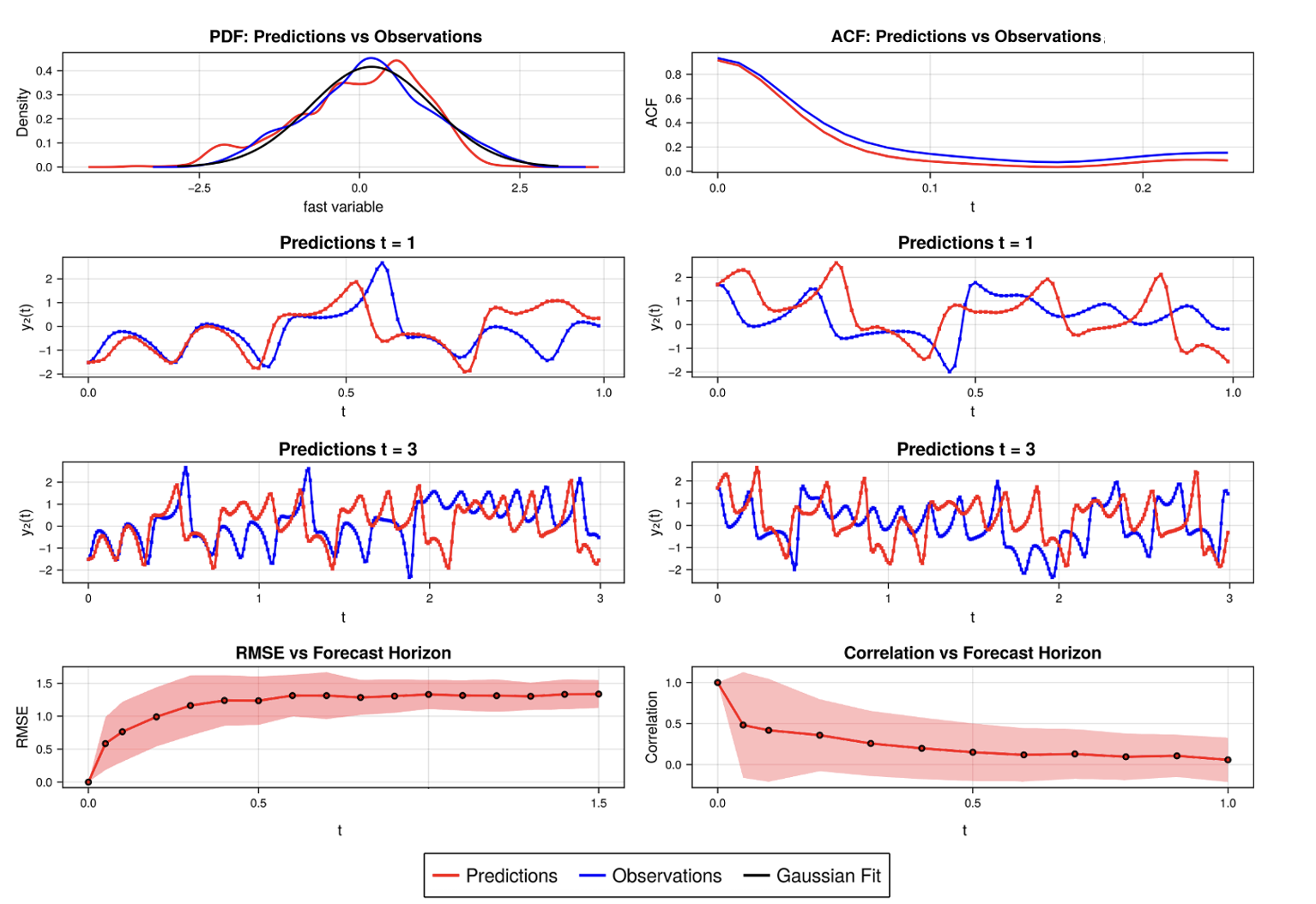}
    \caption{Example III, case a (cycloperiodic signal included in the slow variable)}
    \label{fig:NODEperformance_periodic}
\end{figure}
\subsection{NODE performances for Example IIIb}
\begin{figure}[H]
    \centering
    \includegraphics[width=1\linewidth]{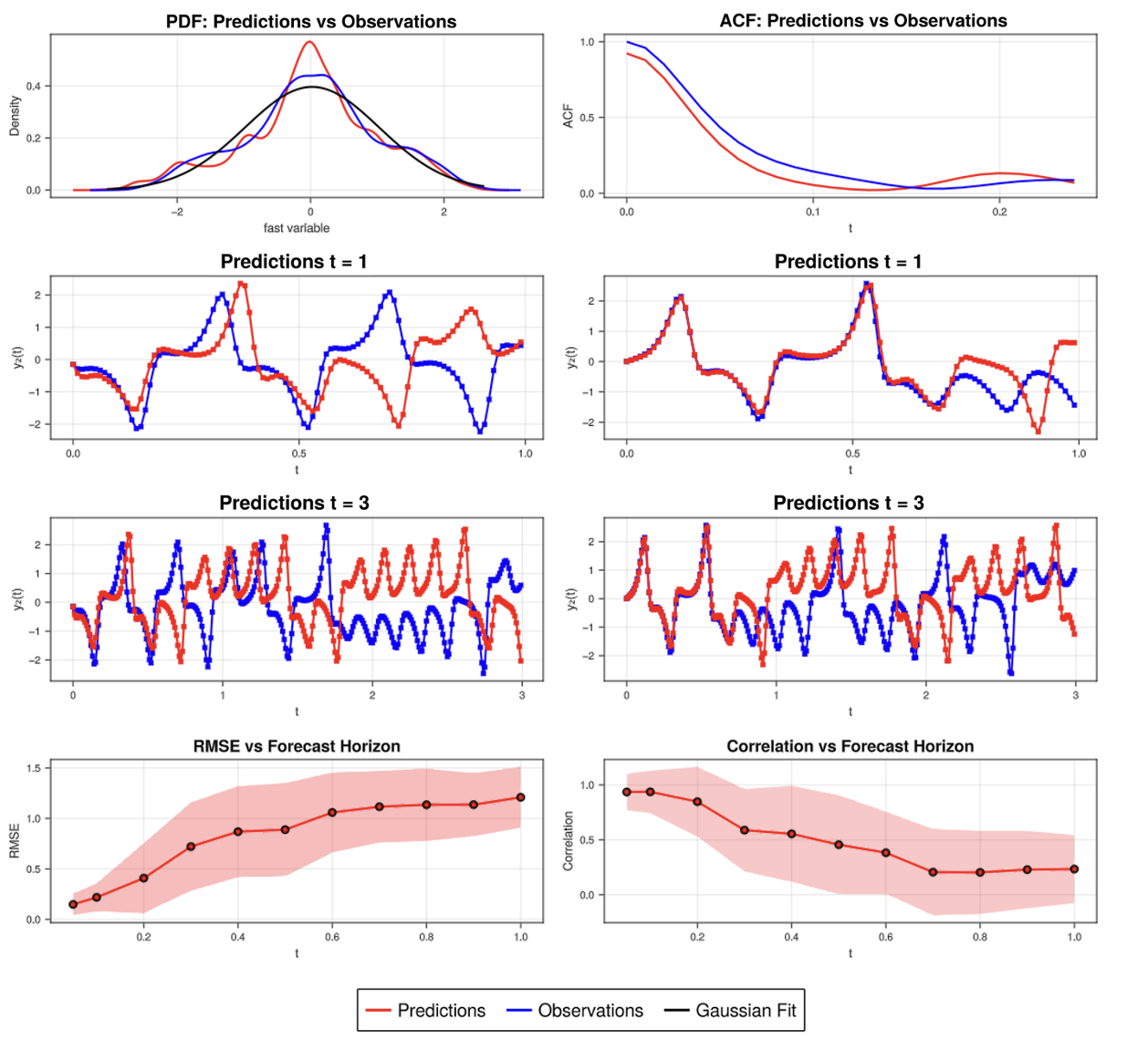}
    \caption{Example III, case b (cycloperiodic signal included in the fast variable)}
    \label{fig:NODEperformancePeriodicFast}
\end{figure}
\newpage
\bibliographystyle{plainnat}
\bibliography{references}

\begin{thebibliography}{57}
\providecommand{\natexlab}[1]{#1}
\providecommand{\url}[1]{\texttt{#1}}
\expandafter\ifx\csname urlstyle\endcsname\relax
  \providecommand{\doi}[1]{doi: #1}\else
  \providecommand{\doi}{doi: \begingroup \urlstyle{rm}\Url}\fi

\bibitem[Baldovin et~al.(2020)Baldovin, Cecconi, and Vulpiani]{baldovin2020understanding}
Marco Baldovin, Fabio Cecconi, and Angelo Vulpiani.
\newblock Understanding causation via correlations and linear response theory.
\newblock \emph{Physical Review Research}, 2\penalty0 (4):\penalty0 043436, 2020.

\bibitem[Berner et~al.(2017)Berner, Achatz, Lott, Imkeller, Franzke, et~al.]{Berner2017StochParam}
Judith Berner, Ulrich Achatz, Francois Lott, Peter Imkeller, Christian L.~E. Franzke, et~al.
\newblock Stochastic parameterization: Toward a new view of weather and climate models.
\newblock \emph{Bulletin of the American Meteorological Society}, 98\penalty0 (3):\penalty0 565--587, 2017.
\newblock \doi{10.1175/BAMS-D-15-00268.1}.

\bibitem[Callaham et~al.(2022)Callaham, Rigas, Loiseau, and Brunton]{callaham2022empirical}
Jared~L Callaham, Georgios Rigas, Jean-Christophe Loiseau, and Steven~L Brunton.
\newblock An empirical mean-field model of symmetry-breaking in a turbulent wake.
\newblock \emph{Science Advances}, 8\penalty0 (19):\penalty0 eabm4786, 2022.

\bibitem[Change(2007)]{change2007climate}
Intergovernmental Panel On~Climate Change.
\newblock Climate change 2007: the physical science basis.
\newblock \emph{Agenda}, 6\penalty0 (07):\penalty0 333, 2007.

\bibitem[Chen et~al.(2018)Chen, Rubanova, Bettencourt, and Duvenaud]{Chen2018Neural}
Ricky T.~Q. Chen, Yulia Rubanova, Jesse Bettencourt, and David~K. Duvenaud.
\newblock Neural ordinary differential equations.
\newblock \emph{Advances in Neural Information Processing Systems}, 31, 2018.

\bibitem[Chorin and Lu(2015)]{chorin2015discrete}
Alexandre~J Chorin and Fei Lu.
\newblock Discrete approach to stochastic parametrization and dimension reduction in nonlinear dynamics.
\newblock \emph{Proceedings of the National Academy of Sciences}, 112\penalty0 (32):\penalty0 9804--9809, 2015.

\bibitem[Cooper and Haynes(2011)]{cooper2011climate}
Fenwick~C Cooper and Peter~H Haynes.
\newblock Climate sensitivity via a nonparametric fluctuation--dissipation theorem.
\newblock \emph{Journal of the Atmospheric Sciences}, 68\penalty0 (5):\penalty0 937--953, 2011.

\bibitem[Falasca et~al.(2025)Falasca, Basinski, Zanna, and Zhao]{falasca2025FDT}
F.~Falasca, A.~Basinski, L.~Zanna, and M.~Zhao.
\newblock {A fluctuation-dissipation theorem perspective on radiative responses to temperature perturbations}.
\newblock \emph{Arxiv (Accepted, in Press in Journal of Climate)}, 2025.
\newblock \doi{https://doi.org/10.48550/arXiv.2408.12585}.

\bibitem[Falasca et~al.(2024)Falasca, Perezhogin, and Zanna]{falasca2024data}
Fabrizio Falasca, Pavel Perezhogin, and Laure Zanna.
\newblock Data-driven dimensionality reduction and causal inference for spatiotemporal climate fields.
\newblock \emph{Physical Review E}, 109\penalty0 (4):\penalty0 044202, 2024.

\bibitem[Fenichel(1979)]{Fenichel1979GSPT}
Neil Fenichel.
\newblock Geometric singular perturbation theory for ordinary differential equations.
\newblock \emph{Journal of Differential Equations}, 31\penalty0 (1):\penalty0 53--98, 1979.
\newblock \doi{10.1016/0022-0396(79)90152-9}.

\bibitem[Ge et~al.(2024)Ge, Nauta, Hagan, and Dinner]{Ge2024PRL}
Hengtai Ge, Nikolas~N. Nauta, Michael~R. Hagan, and Aaron~R. Dinner.
\newblock Data-driven learning of the generalized langevin equation.
\newblock \emph{Physical Review Letters}, 133:\penalty0 077301, 2024.
\newblock \doi{10.1103/PhysRevLett.133.077301}.

\bibitem[Ghil and Lucarini(2020)]{ghil2020physics}
Michael Ghil and Valerio Lucarini.
\newblock The physics of climate variability and climate change.
\newblock \emph{Reviews of Modern Physics}, 92\penalty0 (3):\penalty0 035002, 2020.

\bibitem[Giorgini et~al.(2020)Giorgini, Lim, Moon, and Wettlaufer]{giorgini2020precursors}
Ludovico~T Giorgini, Soon~H Lim, Woosok Moon, and John~S Wettlaufer.
\newblock Precursors to rare events in stochastic resonance.
\newblock \emph{Europhysics letters}, 129\penalty0 (4):\penalty0 40003, 2020.

\bibitem[Giorgini et~al.(2022)Giorgini, Moon, Chen, and Wettlaufer]{giorgini2022non}
Ludovico~T Giorgini, W~Moon, N~Chen, and J~Wettlaufer.
\newblock Non-gaussian stochastic dynamical model for the el niño southern oscillation.
\newblock \emph{Physical Review Research}, 4\penalty0 (2):\penalty0 L022065, 2022.

\bibitem[Giorgini et~al.(2024{\natexlab{a}})Giorgini, Souza, and Schmid]{giorgini2024reduced}
Ludovico~T Giorgini, Andre~N Souza, and Peter~J Schmid.
\newblock Reduced markovian models of dynamical systems.
\newblock \emph{Physica D: Non-linear Phenomena}, 470:\penalty0 134393, 2024{\natexlab{a}}.

\bibitem[Giorgini et~al.(2025{\natexlab{a}})Giorgini, Bischoff, and Souza]{giorgini2025kgmm}
Ludovico~T Giorgini, Tobias Bischoff, and Andre~N Souza.
\newblock Kgmm: A k-means clustering approach to gaussian mixture modeling for score function estimation.
\newblock \emph{arXiv preprint arXiv:2503.18054}, 2025{\natexlab{a}}.

\bibitem[Giorgini et~al.(2025{\natexlab{b}})Giorgini, Bischoff, and Souza]{giorgini2025statistical}
Ludovico~T Giorgini, Tobias Bischoff, and Andre~N Souza.
\newblock Statistical parameter calibration with the generalized fluctuation dissipation theorem and generative modeling.
\newblock \emph{arXiv preprint arXiv:2509.19660}, 2025{\natexlab{b}}.

\bibitem[Giorgini et~al.(2025{\natexlab{c}})Giorgini, Falasca, and Souza]{giorgini2025predicting}
Ludovico~T Giorgini, Fabrizio Falasca, and Andre~N Souza.
\newblock Predicting forced responses of probability distributions via the fluctuation--dissipation theorem and generative modeling.
\newblock \emph{Proceedings of the National Academy of Sciences}, 122\penalty0 (41):\penalty0 e2509578122, 2025{\natexlab{c}}.

\bibitem[Giorgini et~al.(2025{\natexlab{d}})Giorgini, Souza, Lippolis, Cvitanović, and Schmid]{giorgini2025learning}
Ludovico~T Giorgini, Andre~N Souza, Domenico Lippolis, Predrag Cvitanović, and Peter Schmid.
\newblock Learning dissipation and instability fields from chaotic dynamics.
\newblock \emph{arXiv preprint arXiv:2502.03456}, 2025{\natexlab{d}}.

\bibitem[Giorgini(2025)]{giorgini2025data}
Ludovico~Theo Giorgini.
\newblock Data-driven decomposition of conservative and non-conservative dynamics in multiscale systems.
\newblock \emph{arXiv preprint arXiv:2505.01895}, 2025.

\bibitem[Giorgini et~al.(2021)Giorgini, Lim, Moon, Chen, and Wettlaufer]{giorgini2021modeling}
Ludovico~Theo Giorgini, Soon~Hoe Lim, Woosok Moon, Nan Chen, and John~S Wettlaufer.
\newblock Modeling the el nino southern oscillation with neural differential equations.
\newblock In \emph{Thirty-eighth International Conference on Machine Learning (ICML 2021), Time Series Workshop, virtual, July 24, 2021}. International Conference on Machine Learning, 2021.

\bibitem[Giorgini et~al.(2024{\natexlab{b}})Giorgini, Deck, Bischoff, and Souza]{giorgini_response_theory}
Ludovico~Theo Giorgini, Katherine Deck, Tobias Bischoff, and Andre Souza.
\newblock Response theory via generative score modeling.
\newblock \emph{Physical Review Letters}, 133\penalty0 (26):\penalty0 267302, 2024{\natexlab{b}}.

\bibitem[Giorgini et~al.(2025{\natexlab{e}})Giorgini, Bischoff, and Souza]{giorgini2025reduced}
Ludovico~Theo Giorgini, Tobias Bischoff, and Andre~Noguiera Souza.
\newblock Reduced-order modeling of cyclo-stationary time series using score-based generative methods.
\newblock \emph{arXiv preprint arXiv:2508.19448}, 2025{\natexlab{e}}.

\bibitem[H{\"a}nggi et~al.(1990)H{\"a}nggi, Talkner, and Borkovec]{Hanggi1990Kramers}
Peter H{\"a}nggi, Peter Talkner, and Michal Borkovec.
\newblock Reaction-rate theory: fifty years after kramers.
\newblock \emph{Reviews of Modern Physics}, 62\penalty0 (2):\penalty0 251--341, 1990.
\newblock \doi{10.1103/RevModPhys.62.251}.

\bibitem[Hasselmann(1976)]{Hasselmann1976}
K.~Hasselmann.
\newblock Stochastic climate models part i. theory.
\newblock \emph{Tellus}, 28:\penalty0 473--485, 1976.
\newblock \doi{https://doi.org/10.1111/j.2153-3490.1976.tb00696.x}.

\bibitem[Ho et~al.(2020)Ho, Jain, and Abbeel]{Ho2020DDPM}
Jonathan Ho, Ajay Jain, and Pieter Abbeel.
\newblock Denoising diffusion probabilistic models.
\newblock In \emph{Advances in Neural Information Processing Systems 33 (NeurIPS)}, 2020.

\bibitem[Hyv{\"a}rinen and Dayan(2005)]{Hyvarinen2005Score}
Aapo Hyv{\"a}rinen and Peter Dayan.
\newblock Estimation of non-normalized statistical models by score matching.
\newblock \emph{Journal of Machine Learning Research}, 6\penalty0 (4), 2005.

\bibitem[In and Kim(2013)]{in2013introduction}
Francis In and Sangbae Kim.
\newblock \emph{An introduction to wavelet theory in finance: a wavelet multiscale approach}.
\newblock World scientific, 2013.

\bibitem[Jung and H{\"a}nggi(1987)]{JungHanggi1987UCNA}
Peter Jung and Peter H{\"a}nggi.
\newblock Dynamical systems: A unified colored-noise approximation.
\newblock \emph{Physical Review A}, 35\penalty0 (10):\penalty0 4464--4466, 1987.
\newblock \doi{10.1103/PhysRevA.35.4464}.

\bibitem[Keyes et~al.(2023)Keyes, Giorgini, and Wettlaufer]{keyes2023stochastic}
N~D Keyes, L~T Giorgini, and J~S Wettlaufer.
\newblock Stochastic paleoclimatology: Modeling the epica ice core climate records.
\newblock \emph{Chaos}, 33\penalty0 (9):\penalty0 093132, 2023.

\bibitem[Kramers(1940)]{Kramers1940Physica}
H.~A. Kramers.
\newblock Brownian motion in a field of force and the diffusion model of chemical reactions.
\newblock \emph{Physica}, 7\penalty0 (4):\penalty0 284--304, 1940.
\newblock \doi{10.1016/S0031-8914(40)90098-2}.

\bibitem[Kubo(1966)]{Kubo1966FDT}
Ryogo Kubo.
\newblock The fluctuation-dissipation theorem.
\newblock \emph{Reports on Progress in Physics}, 29\penalty0 (1):\penalty0 255--284, 1966.
\newblock \doi{10.1088/0034-4885/29/1/306}.

\bibitem[Lei et~al.(2016)Lei, Baker, and Li]{Lei2016GLE}
Huan Lei, Nathan~A. Baker, and Xiantao Li.
\newblock Data-driven parameterization of the generalized langevin equation.
\newblock \emph{Proceedings of the National Academy of Sciences}, 113\penalty0 (50):\penalty0 14183--14188, 2016.
\newblock \doi{10.1073/pnas.1609587113}.

\bibitem[Lim et~al.(2020)Lim, Theo~Giorgini, Moon, and Wettlaufer]{lim2020predicting}
Soon~Hoe Lim, Ludovico Theo~Giorgini, Woosok Moon, and John~S Wettlaufer.
\newblock Predicting critical transitions in multiscale dynamical systems using reservoir computing.
\newblock \emph{Chaos: An Interdisciplinary Journal of Nonlinear Science}, 30\penalty0 (12), 2020.

\bibitem[Lorenz(2017)]{lorenz2017deterministic}
Edward~N Lorenz.
\newblock Deterministic nonperiodic flow 1.
\newblock In \emph{Universality in Chaos, 2nd edition}, pages 367--378. Routledge, 2017.

\bibitem[Lucarini and Chekroun(2023)]{LucariniChekroun2023}
V.~Lucarini and M.D. Chekroun.
\newblock {Theoretical tools for understanding the climate crisis from Hasselmann’s programme and beyond}.
\newblock \emph{Nat Rev Phys (2023)}, 2023.
\newblock URL \url{https://doi.org/10.1038/s42254-023-00650-8}.

\bibitem[Majda et~al.(2008)Majda, Franzke, and Khouider]{majda_fdt}
Andrew~J Majda, Christian Franzke, and Boualem Khouider.
\newblock An applied mathematics perspective on stochastic modelling for climate.
\newblock \emph{Philosophical Transactions of the Royal Society A: Mathematical, Physical and Engineering Sciences}, 366\penalty0 (1875):\penalty0 2427--2453, 2008.

\bibitem[Marconi et~al.(2008)Marconi, Puglisi, Rondoni, and Vulpiani]{marconi2008}
U.~M.~B. Marconi, A.~Puglisi, L.~Rondoni, and A.~Vulpiani.
\newblock Fluctuation-dissipation: Response theory in statistical physics.
\newblock \emph{Physics Reports}, 461\penalty0 (4-6):\penalty0 111--195, 2008.
\newblock \doi{10.1016/j.physrep.2008.02.002}.

\bibitem[Meier-Schellersheim et~al.(2009)Meier-Schellersheim, Fraser, and Klauschen]{meier2009multiscale}
Martin Meier-Schellersheim, Iain~DC Fraser, and Frederick Klauschen.
\newblock Multiscale modeling for biologists.
\newblock \emph{Wiley Interdisciplinary Reviews: Systems Biology and Medicine}, 1\penalty0 (1):\penalty0 4--14, 2009.

\bibitem[Mori(1965)]{Mori1965Transport}
Hazime Mori.
\newblock Transport, collective motion, and brownian motion.
\newblock \emph{Progress of Theoretical Physics}, 33\penalty0 (3):\penalty0 423--455, 1965.
\newblock \doi{10.1143/PTP.33.423}.

\bibitem[Pavliotis and Stuart(2008)]{pavliotis2008multiscale}
Grigorios~A Pavliotis and Andrew Stuart.
\newblock \emph{Multiscale methods: averaging and homogenization}, volume~53.
\newblock Springer Science \& Business Media, 2008.

\bibitem[Platt et~al.(2023)Platt, Penny, Smith, Chen, and Abarbanel]{platt2023constraining}
Jason~A Platt, Stephen~G Penny, Timothy~A Smith, Tse-Chun Chen, and Henry~DI Abarbanel.
\newblock Constraining chaos: Enforcing dynamical invariants in the training of reservoir computers.
\newblock \emph{Chaos: An Interdisciplinary Journal of Nonlinear Science}, 33\penalty0 (10), 2023.

\bibitem[Rackauckas et~al.(2020)Rackauckas, Ma, Martensen, Warner, Zubov, Supekar, Skinner, Ramadhan, and Edelman]{Rackauckas2020UDE}
Christopher Rackauckas, Yingbo Ma, Julius Martensen, Collin Warner, Kirill Zubov, Rohit Supekar, Dominic Skinner, Ali Ramadhan, and Alan Edelman.
\newblock Universal differential equations for scientific machine learning.
\newblock \emph{Nature Communications}, 11:\penalty0 3621, 2020.
\newblock \doi{10.1038/s41467-020-17265-8}.

\bibitem[Ruelle(1998)]{Ruelle1998GLRF}
David Ruelle.
\newblock General linear response formula in statistical mechanics and the fluctuation-dissipation theorem far from equilibrium.
\newblock \emph{Physics Letters A}, 245\penalty0 (3-4):\penalty0 220--224, 1998.
\newblock \doi{10.1016/S0375-9601(98)00419-8}.

\bibitem[Ruelle(2009)]{Ruelle2009Review}
David Ruelle.
\newblock A review of linear response theory for general differentiable dynamical systems.
\newblock \emph{Nonlinearity}, 22\penalty0 (4):\penalty0 855--885, 2009.
\newblock \doi{10.1088/0951-7715/22/4/011}.

\bibitem[Schneider et~al.(2017)Schneider, Lan, Stuart, and Teixeira]{schneider2017earth}
Tapio Schneider, Shiwei Lan, Andrew Stuart, and Joao Teixeira.
\newblock Earth system modeling 2.0: A blueprint for models that learn from observations and targeted high-resolution simulations.
\newblock \emph{Geophysical Research Letters}, 44\penalty0 (24):\penalty0 12--396, 2017.

\bibitem[Song and Ermon(2019)]{SongErmon2019NCSN}
Yang Song and Stefano Ermon.
\newblock Generative modeling by estimating gradients of the data distribution.
\newblock In \emph{Advances in Neural Information Processing Systems 32 (NeurIPS)}, 2019.

\bibitem[Song et~al.(2021)Song, Sohl-Dickstein, Kingma, Kumar, Ermon, and Poole]{song_sde}
Yang Song, Jascha Sohl-Dickstein, Diederik~P Kingma, Abhishek Kumar, Stefano Ermon, and Ben Poole.
\newblock Score-based generative modeling through stochastic differential equations.
\newblock \emph{arXiv preprint arXiv:2011.13456}, 2021.

\bibitem[Souza(2024{\natexlab{a}})]{souza2024representing_a}
Andre~N Souza.
\newblock Representing turbulent statistics with partitions of state space. part 1. theory and methodology.
\newblock \emph{Journal of Fluid Mechanics}, 997:\penalty0 A1, 2024{\natexlab{a}}.

\bibitem[Souza(2024{\natexlab{b}})]{souza2024representing_b}
Andre~N Souza.
\newblock Representing turbulent statistics with partitions of state space. part 2. the compressible euler equations.
\newblock \emph{Journal of Fluid Mechanics}, 997:\penalty0 A2, 2024{\natexlab{b}}.

\bibitem[Souza and Silvestri(2024)]{souza2024modified}
Andre~N. Souza and Simone Silvestri.
\newblock A modified bisecting k-means for approximating transfer operators: Application to the lorenz equations.
\newblock \emph{arXiv preprint arXiv:2412.03734}, 2024.
\newblock \doi{10.48550/arXiv.2412.03734}.
\newblock URL \url{https://doi.org/10.48550/arXiv.2412.03734}.
\newblock Submitted to arXiv on Dec 4, 2024.

\bibitem[Steinhauser(2017)]{steinhauser2017computational}
Martin~Oliver Steinhauser.
\newblock \emph{Computational multiscale modeling of fluids and solids}.
\newblock Springer, 2017.

\bibitem[Takens(1981)]{Takens}
F.~Takens.
\newblock \emph{{Detecting strange attractors in turbulence, in Dynamical Systems and Turbulence}}, volume 898, page 21–48.
\newblock Springer, Berlin,Heidelberg, 1981.

\bibitem[Uhlenbeck and Ornstein(1930)]{UhlenbeckOrnstein1930}
G.~E. Uhlenbeck and L.~S. Ornstein.
\newblock On the theory of the brownian motion.
\newblock \emph{Physical Review}, 36\penalty0 (5):\penalty0 823--841, 1930.
\newblock \doi{10.1103/PhysRev.36.823}.

\bibitem[Vincent(2011)]{Vincent2011DAE}
Pascal Vincent.
\newblock A connection between score matching and denoising autoencoders.
\newblock \emph{Neural Computation}, 23\penalty0 (7):\penalty0 1661--1674, 2011.
\newblock \doi{10.1162/NECO_a_00142}.

\bibitem[Vroylandt et~al.(2022)Vroylandt, Gouden{\`e}ge, Monmarch{\'e}, Pietrucci, and Rotenberg]{Vroylandt2022LLnonMarkov}
Hadrien Vroylandt, Ludovic Gouden{\`e}ge, Pierre Monmarch{\'e}, Fabio Pietrucci, and Benjamin Rotenberg.
\newblock Likelihood-based non-markovian models from molecular dynamics.
\newblock \emph{Proceedings of the National Academy of Sciences}, 119\penalty0 (13):\penalty0 e2117586119, 2022.
\newblock \doi{10.1073/pnas.2117586119}.

\bibitem[Zwanzig(1961)]{Zwanzig1961Memory}
Robert Zwanzig.
\newblock Memory effects in irreversible thermodynamics.
\newblock \emph{Physical Review}, 124\penalty0 (4):\penalty0 983--992, 1961.
\newblock \doi{10.1103/PhysRev.124.983}.

\end{thebibliography}
\end{document}